\documentclass[11pt]{article}
\usepackage[margin=1in]{geometry}
\usepackage[T1]{fontenc}
\usepackage{lmodern}
\usepackage{microtype}
\usepackage{amsmath,amssymb,amsthm}
\usepackage{booktabs}
\usepackage{graphicx}
\usepackage{xcolor}
\usepackage[numbers,sort&compress]{natbib}
\usepackage[hidelinks]{hyperref}
\usepackage{enumitem}
\usepackage{tikz}
\usetikzlibrary{positioning,arrows.meta}
\setlist{nosep,leftmargin=1.4em}

\hypersetup{pdftitle={A Cost That Practice Accuracy Hides: Worked Solutions in Self-Generated Training Data and the Compositional Use of New Rules},pdfauthor={Jiantong Xu}}

\newtheorem{proposition}{Proposition}
\newcommand{\pp}{\,pp}
\newcommand{\ci}[2]{[#1,\,#2]}
\newcommand{\genT}{\textsc{Gen}}
\newcommand{\instT}{\textsc{Inst}}
\newcommand{\ansT}{\textsc{Ans}}
\newcommand{\keptT}{\textsc{Kept}}
\definecolor{genblue}{HTML}{2A78D6}
\definecolor{instorange}{HTML}{EB6834}

\title{A Cost That Practice Accuracy Hides:\\
Worked Solutions in Self-Generated Training Data\\
and the Compositional Use of New Rules}
\author{Jiantong Xu\thanks{ORCID: \href{https://orcid.org/0009-0005-7470-9236}{https://orcid.org/0009-0005-7470-9236}.}\\
Independent Researcher}
\date{}

\begin{document}
\maketitle

\begin{abstract}
A worked solution can make self-generated training targets shorter and cheaper, but accuracy on the practiced operation may hide how well a model can reuse newly learned rules.
We study this choice in a 2B-parameter reasoning model learning two invented quantitative rules.
The model generates targets from a rule sheet alone or from the sheet plus a program-written worked solution; targets are accepted for final-answer correctness, and the support is removed for fine-tuning and evaluation.
Training questions request forward use in one segment; composition requires sequential reuse of the same rule across two segments.
After six passes, using a program-checked numerical signature, sheet-only targets lead by 24.3 percentage points (95\% item-bootstrap CI 14.6--34.0) on composition in a prospective test.
On unseen argument values, observed forward accuracy is identical (87.5\% each), but composition accuracy is 58.3\% versus 23.6\%, a 34.7-point gap (25.7--43.8).
An independently sampled second target-bank pair repeats the direction (+14.6 points, 2.1--26.0).
Ten passes on the shorter worked-solution targets, approximately matching the domain tokens of six sheet-only passes while changing other training quantities, narrow the gap to 16.3 points (6.3--26.7) in the first bank pair and 6.2 points ($-6.3$ to 18.8) in the second, where the pre-specified test does not hold.
At six passes each, the worked-solution policy uses fewer capture tokens and less update time, yet the trained models use 2.6 times as many generated tokens per composition success under the numerical signature (2.1 times in the second bank pair).
Choosing a capture context therefore requires testing how the new rules are reused, alongside the inference cost of those uses, rather than relying on forward accuracy alone.

\end{abstract}

\section{Introduction}
\label{sec:intro}

Teams that deploy language models often need them to apply rules the models have never seen: the equations of a new battery-cell model, a revised pricing formula, a newly issued design standard.
Putting the rules in every prompt costs context and latency and fails when the rules are long or the model must work without them, so the rules are put into the weights.
One established recipe lets the model write its own training data \citep{snell2022context,padmanabhan2023propagating,park2025newnews,yang2025rule}: the model answers practice questions with the rules in its context, outputs with a correct final answer are kept, and the model is fine-tuned on them \emph{without} the rules in the input.
We call the material in the context while the model writes a training target the \emph{capture context}.

What the capture context should contain is an active design choice, and supplying a worked solution or a verified answer alongside the rules is a well-motivated strategy.
Several self-training methods show the target writer an answer or a solution that the trained model will not see: STaR uses the correct answer as a hint for rationales of problems that the model failed \citep{zelikman2022star}, SCORE includes a correct solution of the same question when a small model writes critiques of its own wrong solutions \citep{zhang2024score}, and on-policy self-distillation (OPSD) conditions the teacher on a verified reference solution while the student sees only the question \citep{zhao2026opsd}.
Most directly, in the knowledge-acquisition ablation of Self-Distillation Fine-Tuning (SDFT), an on-policy self-teacher given both a source article and a worked answer reached 89\% strict accuracy on new facts, compared with 75\% given the article alone \citep[App.~A.2]{shenfeld2026sdft}.
A solver, a reference program or a stronger model can often supply a worked solution, and the targets written with it can be shorter and cheaper to train on \citep[Sec.~4]{kim2026selfdistillation}.
Candidate contexts are then judged by what a data pipeline can measure cheaply: the cost of capture and training, and accuracy on held-out questions like the practice questions.
The question of this paper is whether that criterion is enough.

Yet the benefits of solution-assisted supervision are not universal.
Fine-tuning offline on answer-conditioned \citep{lee2026answerconditioned} or solution-guided \citep[Sec.~4]{kim2026selfdistillation} reasoning traces can degrade mathematical reasoning even when every training trace passes a correctness filter; in on-policy self-distillation, the best privileged context changes with model scale and task \citep{tian2026selfteacher}, privileged solution views add little beyond reference-free training \citep{zhang2026ample}, and a solution of another problem can do better than the correct reference \citep{shrestha2026rethinking}.
These findings point to a tension: \emph{a training target that is easy to generate and verify is not necessarily one that teaches the model to use the underlying knowledge on its own.}
The studies differ in procedure, in whether the knowledge is new, and in what they measure.
Work on composable knowledge editing and synthetic function acquisition already connects context-conditioned learning with the use of new knowledge \citep{liu2026hpse,liu2026mixsd}. Our question is narrower: when the rule sheet and training procedure are shared, can adding a worked solution during offline target generation leave forward accuracy similar but change sequential reuse of the new rules?
Machine learning has a name for the general risk: a pipeline is \emph{underspecified} when it can return predictors that are equivalent on held-out data from the training distribution but behave differently elsewhere \citep{damour2020underspecification}.
We ask whether one common choice in self-generated training data, a worked solution in the capture context, is such a case, and how large the hidden difference is.
In a setting built to measure it, it is: two capture contexts with nearly the same forward accuracy give models whose composition of the new rules differs by 24 to 35 points, and the context that is cheaper to train uses more generated tokens per composition success under the numerical value signature.

\begin{figure}[t]
\centering
\begin{tikzpicture}[
  font=\footnotesize,
  box/.style={draw=black!55, rounded corners=2pt, align=left, inner sep=4pt, text width=#1},
  arrow/.style={-{Stealth[length=2.2mm]}, thick, black!70}
]
\node[anchor=north west, font=\small] (lab) at (0,0) {(a)};
\node[box=5.75cm, anchor=north west] (cap) at (0.55,0) {
  \textbf{1. Capture} (base model)\\[2pt]
  context: one of the two below\\
  question: a forward practice question\\
  kept if the final answer is correct
};
\node[box=5.75cm, below=0.12cm of cap, fill=genblue!12] (gen) {\genT: the rule sheet};
\node[box=5.75cm, below=0.08cm of gen, fill=instorange!14] (inst) {\instT: the sheet + a program-written worked solution};
\node[box=3.6cm, right=0.5cm of cap.north east, anchor=north west] (train) {
  \textbf{2. Train} (LoRA+)\\[2pt]
  input: the question only\\
  target: the kept answer\\
  6 or 10 passes over\\
  128 questions
};
\node[box=4.0cm, right=0.5cm of train.north east, anchor=north west] (test) {
  \textbf{3. Test} (no context)\\[2pt]
  \begin{tabular}{@{}l@{\quad}l@{}}
  forward use & practiced \\
  inversion & not requested \\
  composition & not requested
  \end{tabular}
};
\draw[arrow] (cap.east |- train.west) -- (train.west);
\draw[arrow] (train.east) -- (test.west |- train.east);
\end{tikzpicture}

\vspace{2pt}
\makebox[\linewidth][l]{\small (b)}

\smallskip
\includegraphics[width=\linewidth]{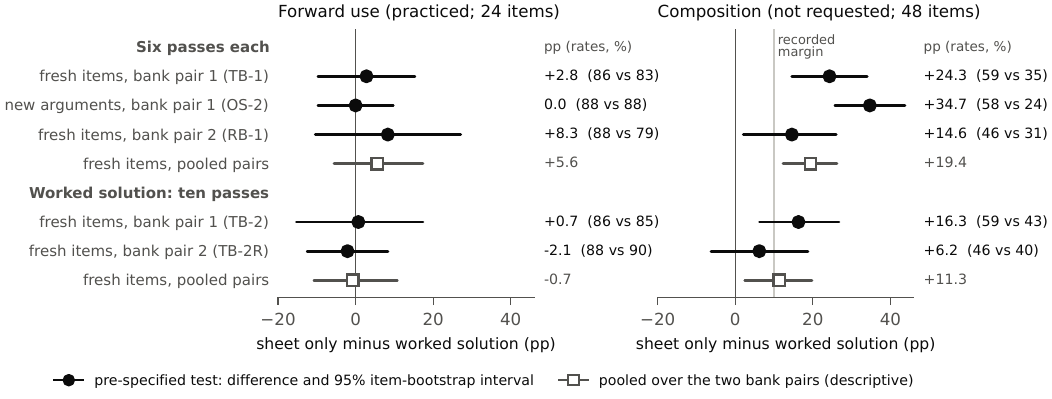}
\caption{\textbf{Forward accuracy does not reveal the capture context; composition does.}
(a)~The two capture policies differ only in the capture context. Completions accepted for final-answer correctness become training targets, and the model is trained on them with the context removed.
(b)~Success without any context, sheet-only (\genT) minus worked-solution (\instT) policy, in percentage points with 95\% item-bootstrap intervals (value signature; rounded success rates in parentheses). Forward differences are small and their intervals contain zero; composition differences are positive in every row. Training the worked-solution policy to about the same number of domain tokens (ten passes) shrinks the composition difference, and with the second bank pair the recorded test (TB-2R) does not hold. Bank pair~2 was not evaluated on new argument values. Pooled rows average the two bank pairs per item and are descriptive.}
\label{fig:story}
\end{figure}
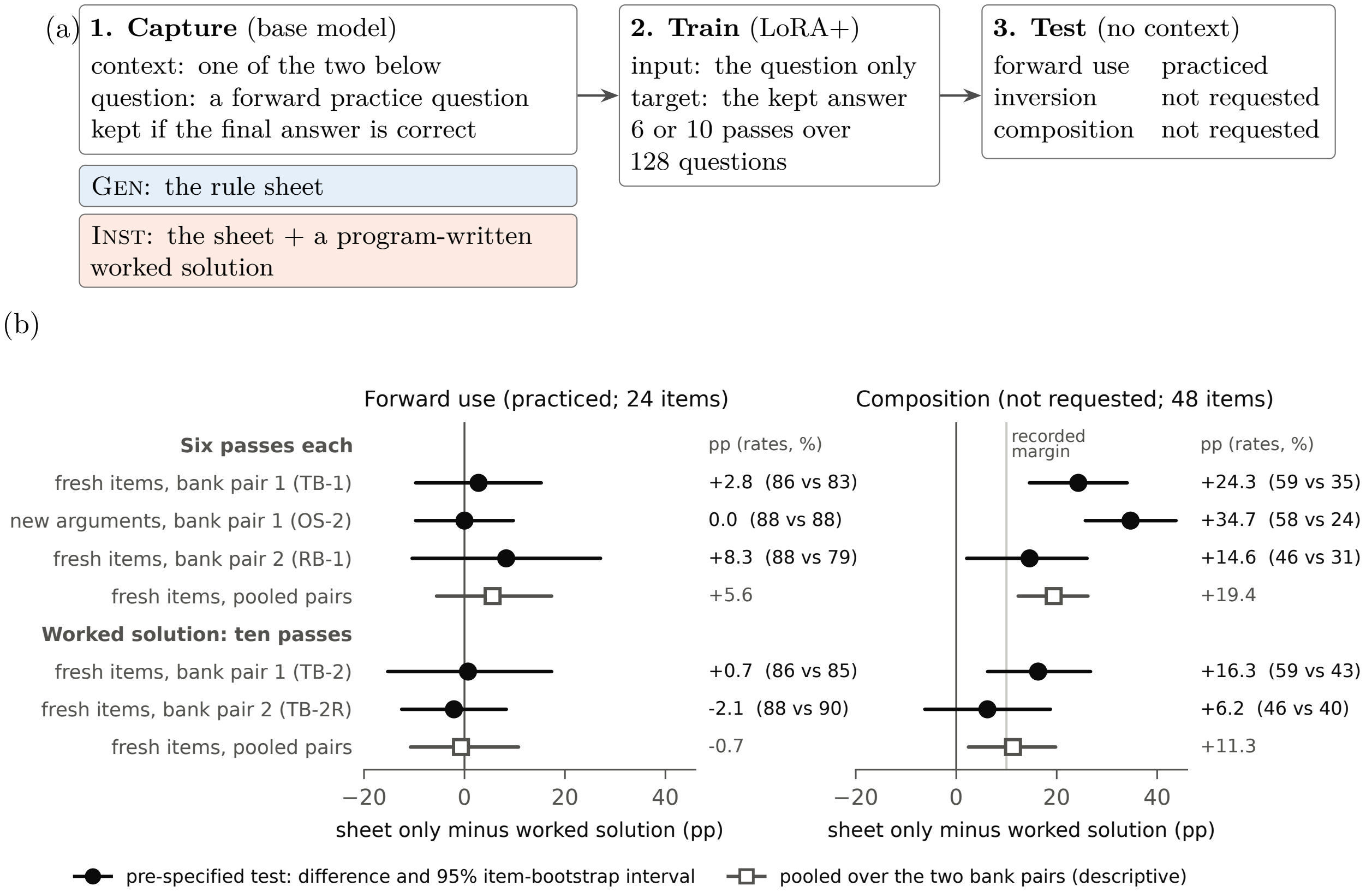

We study the question in a domain built for it (Section~\ref{sec:setting}): a fictional battery cell with two newly assigned quantitative rules and exact program keys.
A 2B-parameter reasoning model writes its own targets for 128 forward questions, each about one segment of constant current, under one of two capture contexts: the rule sheet (\genT), or the same sheet plus a worked solution that a short program computes from the sheet and the question (\instT).
The worked solution adds no rule specification that the sheet lacks, the model can answer the practice questions from the sheet alone, and the instruction is the same in both contexts.
Both policies share the acceptance rule, the target selection, the training recipe and the evaluation conditions; what differs is the targets that each context produces, and with them the target lengths and the numbers of supervised tokens and updates per pass.
The trained models answer without any context, on the practiced forward operation and on two operations that no training question requests: inversion, which recovers a rule parameter from an observed value, and composition, which applies a rule over two consecutive segments, so that the state at the end of the first segment is the start of the second.
Composition here is the sequential reuse of one learned rule, not a combination of the two rules.
We compare the two \emph{complete policies}; we do not isolate which property of the targets is responsible (Section~\ref{sec:identifies}). Unless labeled otherwise, the results below use the value signature, a program check for the keyed numerical values (Section~\ref{sec:setting}).

\paragraph{Findings.}
Each experiment answers one question about the hidden difference (Figure~\ref{fig:story}).
\emph{Does practice-like accuracy show how well each policy can use the rules?}
No: after six passes both policies reach high forward accuracy, but their composition differs. We found the difference after the fact on the first evaluation panel and confirmed it prospectively on new items with a new training seed, $+24.3$ points \ci{14.6}{34.0} (Section~\ref{sec:main}).
\emph{Is the difference recall of the trained argument values?}
No: on items whose rule arguments occur in no training question, forward accuracy stayed identical (87.5\% each), while composition success was 58.3\% for sheet-only and 23.6\% for worked-solution targets, a difference of $+34.7$ points \ci{25.7}{43.8}, as the configuration-bound form of our account had predicted before the test (Section~\ref{sec:offsup}).
\emph{How much of the gap is explained by training amount?}
Part of it. Worked-solution targets are shorter, so we trained that policy for ten passes, which brings its domain tokens close to those of six sheet-only passes but also adds replay, updates and schedule cycles; with the original target banks it narrowed the gap but still trailed by 16.3 points \ci{6.3}{26.7} (Section~\ref{sec:dosebundle}).
\emph{Does the effect depend on a single sample of training targets?}
A second, independently captured bank pair repeated the direction at six passes, $+14.6$ points \ci{2.1}{26.0}; at matched dose its difference was $+6.2$ points \ci{-6.3}{18.8}, and the pre-specified test did not hold, so the size of the gap at matched dose is uncertain (Section~\ref{sec:banks}).
\emph{Does a supplied final answer alone produce the cost?}
We cannot tell: an answer-only context failed its forward-accuracy gate twice, and the reading at its fixed second look is weak (Section~\ref{sec:ans}).

\paragraph{What this means for practice.}
In the numbers that a data pipeline usually logs, the worked solution looks efficient: with the first bank pair, 42\% fewer generated tokens per accepted target and 28\% less update time for six passes, with nearly the same acceptance rate and forward accuracy.
At deployment, its six-pass models have fewer value-signature successes on composition and spend 2.6 times as many generated tokens per success (2.1 times with the second bank pair; Section~\ref{sec:costs}).
The ten-pass extension narrows the accuracy difference but removes the measured update-time saving.
Capture contexts should therefore be compared on operations that the training data do not request before one is chosen.

\paragraph{An account with a prospective test.}
An idealized gradient-flow surrogate illustrates how equal practice accuracy and divergent transfer can coexist: the training targets set the direction of parameter updates, whereas practice accuracy tracks progress along that direction rather than the direction itself (Section~\ref{sec:account}).
It rests on stated assumptions, including observation links shared by the two policies, predicts orderings rather than ratios, and does not identify a mechanism of the network.
Before the new-argument test ran, we recorded what each form of the account and a value-association alternative predicted; the configuration-bound form was the one that held.
Two later predictions failed: the dose-matched second bank pair did not keep the recorded margin, and the answer-only policy did not lie near the sheet-only policy (Section~\ref{sec:tests}).

\paragraph{Contributions.}
The paper makes two main contributions.
(i)~A controlled demonstration that similar observed forward accuracy can conceal large differences in the sequential reuse of newly assigned rules. Prospective tests, new argument values, independently sampled targets and dose extensions establish both the separation and its training-dose boundary.
(ii)~A measurement of the associated cost trade-off: at six passes per policy, savings in capture tokens and update time accompany 2.6 times (second bank pair: 2.1 times) as many generated tokens per composition success under the value signature. The ten-pass extension reduces the accuracy gap but costs more update time than six sheet-only passes.
The evaluation construction and the gradient-flow surrogate support these findings. Exact keys and separate forward, inversion and composition items make the comparison reusable; the surrogate supplies an explicit account with its full prediction record, including the failed predictions, in Sections~\ref{sec:account} and~\ref{sec:tests}.
The comparison is between complete policies in one invented domain, with one main model and one training recipe; which property of the targets carries the difference, and how fully each policy has mastered the rules, remain open (Section~\ref{sec:limits}).

\section{Related work and what is new here}
\label{sec:related}

\paragraph{Answers and worked solutions when training targets are written.}
Learning with privileged information gives a teacher information that the student will not have at test time; generalized distillation trains the student on the outputs of such a teacher \citep{lopezpaz2015unifying}.
Context distillation is the language-model form: a model with extra context in its prompt writes targets for itself without that context \citep{snell2022context}.
Our capture context is privileged information of this kind, available only while targets are written.
Answers and solutions are common forms of this information.
STaR's rationalization step uses the correct answer as a hint for a rationale that is then trained without it \citep{zelikman2022star}; SCORE puts a correct solution of the same question into the prompt as a hint when a small model critiques its own wrong solutions, and removes the hint for fine-tuning \citep[Sec.~3.1]{zhang2024score}; and OPSD conditions the teacher on a verified reference solution while the student, which samples its own trajectories, sees only the question \citep[Sec.~3.2]{zhao2026opsd}.
In the knowledge-acquisition ablation of SDFT, an on-policy teacher that sees a source article and a worked answer gives 89\% strict accuracy on new facts, against 75\% with the article alone \citep[App.~A.2]{shenfeld2026sdft}.

\paragraph{When the added information teaches less.}
\citet{lee2026answerconditioned} hold the generator, problems, correctness filter and fine-tuning recipe fixed and vary only whether the generator sees the gold answer with a request to reach it: training Qwen3-8B on its own answer-conditioned chains instead of answer-blind ones lowers MATH-500 accuracy by 16.2 points, chain length does not explain the loss, and with the answer visible but a neutral instruction the loss falls to about one point (their Sections~3 and~4.1).
They also report a train--test length inversion: the answer-conditioned training chains are shorter, yet the trained student writes longer chains at test time while scoring worse (their Section~4.1, measured in characters).
\citet[Sec.~4]{kim2026selfdistillation} fine-tune offline on 800 correct responses written with or without a full solution in the context; the solution-guided responses are about 2k instead of 12k tokens long, and training on them degrades every mathematics benchmark they report, which they attribute to suppressed expressions of uncertainty in targets that presuppose information unavailable at inference.
\citet{tian2026selfteacher} compare five granularities of privileged context for an on-policy self-teacher, in domain and on four transfer benchmarks, and find that the preferred context changes with scale and task and that the teacher--student discrepancy at initialization does not order the contexts by utility (their Section~4.3).
\citet{zhang2026ample} hold the verified answer fixed across six privileged views, from the answer alone to the full solution trace, and find that reference-free on-policy self-distillation accounts for much of the improvement; the added value of a view is small and does not grow with how much of the solution it reveals (their Sections~2.3 and~3.1).
\citet{penaloza2026pidistill} distil privileged information into agent policies, relate success to the divergence that the information induces and to its utility (their Section~7), and detect references to the privileged information in teacher traces with a keyword list (their Appendix~D.1).
\citet{shrestha2026rethinking} find that in on-policy self-distillation the correct reference does not give a consistent benefit and that a solution of another problem can do better.
Other methods replace full solutions with structural guidance or skill cards \citep{zhao2026psopsd,jiang2026skald}.
Together, these studies establish that what the target writer sees matters and that solution-guided offline targets can harm reasoning.

\paragraph{New knowledge, its use and its composition.}
For new knowledge, \citet{padmanabhan2023propagating} distil entity definitions, \citet{akyurek2024deductive} train on self-generated implications, \citet{park2025newnews} compare self-generated practice formats for hypothetical news, including new mathematical operations, and show that keeping the news in the training input impairs learning (contextual shadowing, which our kept-sheet control replicates), and \citet{yang2025rule} distil a textual rule and test a new operator on larger operands.
HPSE is a close precedent: it uses self-distillation from a privileged in-context state to improve the composability of unstructured knowledge edits, including multi-hop use \citep{liu2026hpse}.
Its intervention is a hybrid rollout that supplies missing facts where the student's trajectory lacks coverage (their Sections~2 and~3).
MixSD constructs supervision by mixing tokens from knowledge-conditioned and prior-conditioned distributions of the base model \citep{liu2026mixsd}.
Its KGFunc evaluation tests synthetic arithmetic operations on new inputs; a separate unseen-operation split probes retained arithmetic capabilities, with few-shot input--output examples still supplied (their Sections~3--5).
These studies already connect context-conditioned learning with the use and generalization of new knowledge.
\citet{press2022compositionality} measure how often a model answers every sub-question but not the composed question; \citet{yuan2025composing} separate atomic string functions from their composition and study what training makes composition generalize; \citet{tang2026opd} find that on-policy distillation transfers compositional skill but little new factual knowledge; and \citet{allenzhu2023physics31} show that stored knowledge need not be extractable in another format.
\citet{gekhman2024newknowledge} relate fine-tuning on new knowledge to hallucination; we observe a related loss of abstention.

\paragraph{Similar on held-out practice, different elsewhere.}
An underspecified pipeline returns predictors with equivalent held-out performance in the training domain that behave differently in deployment \citep{damour2020underspecification}, and shortcut learning names decision rules that do well on standard benchmarks but fail to transfer to harder test conditions \citep{geirhos2020shortcut}.
Our comparison makes this evaluation concern concrete: changing the capture context produces trained states with similar observed forward accuracy but different compositional performance.
At six passes, every paired training seed on fresh items favors the sheet-only policy. The question is what an observed forward score reveals, not whether the policies have identical latent mastery.

\paragraph{What is new here.}
Our contribution is a controlled test of a particular offline data-generation choice: whether adding a worked solution to the same rule sheet changes the model's later sequential reuse of newly assigned rules, even when forward evaluation reports similar accuracy.
HPSE changes rollout coverage, and MixSD changes token-level mixing; here the intervention is the presence of a worked solution while a fixed bank of training targets is generated.
The instruction, acceptance rule, target selection, training recipe and evaluation conditions are shared, while the generated targets and their associated training exposure can differ.
Unlike the answer-conditioned prompts of \citet{lee2026answerconditioned}, our instruction does not ask the model to reach a shown answer; this is a different prompting condition, not evidence of a different mechanism.
The experiment separates practiced forward use from unrequested inversion and sequential reuse, including reuse at new rule-argument values.
The composition gap holds in a prospective test, at new argument values and with an independently sampled bank pair; longer training narrows it, and the matched-dose test holds for the original banks but not the second pair.
We quantify the accompanying capture-token, update-time and inference-token trade-offs, with inference cost measured per value-signature success.
The train--test length inversion reported by \citet{lee2026answerconditioned} appears here on composition but not on the practiced forward operation (Section~\ref{sec:diagnostics}).
Because our study examines offline target capture, it complements rather than adjudicates on-policy self-distillation studies.
Our literature search was focused, not exhaustive (Appendix~\ref{app:literature}).

\section{Experimental design}
\label{sec:setting}

\subsection{New rules with exact keys}

The \emph{Kestrel cell} is a fictional battery element with two rules (Appendix~\ref{app:sheet} gives the full sheet).
\textbf{T1} gives the voltage $v$ of a ``Varn layer'' a time $h$ after the start of a constant-current segment: $v = r\,v_\text{start} + (1-r)\,gI$ with $r = (1+h/\tau)^{-2}$, where $\tau$ is a time constant and $g$ a gain; $v$ is continuous at a current step.
\textbf{T2} gives the usable charge at a discharge current $I$: $Q_\text{use} = Q_\text{nom}\,(1 - k\sqrt{I/I_\text{ref}})$; a segment lowers the state of charge (SOC) by $Ih/Q_\text{use}$.
Both rules resemble familiar battery models but differ from them: first-order relaxation is exponential, and a simple charge model has a constant capacity.
The rules are newly assigned to a fictional cell, and the base model never produces a Kestrel value without the sheet (Section~\ref{sec:dose}), although the mathematical forms are ordinary.
A seeded program generates every item and computes the exact Kestrel answer and the answers that the familiar alternatives imply; it keeps an item only if these values are separated by a fixed margin.

\begin{figure}[tbp]
\small
\fbox{\begin{minipage}{0.97\linewidth}
\textbf{Forward (practiced), A-T2.} \emph{The cell has $Q_\text{nom}$ = 15 Ah, $k$ = 1/6, and $I_\text{ref}$ = 4 A. At time 0, its SOC is 4/5. A constant discharge current $I$ = 16 A flows from time 0 to time 900 s. Calculate the usable charge $Q_\text{use}$ at this current. Calculate the SOC at the end of the discharge.}
Key: $Q_\text{use} = 15(1 - \tfrac16\sqrt{16/4}) = 10$ Ah; final SOC $= 4/5 - 4/10 = 2/5$. (Constant capacity would give 0.533.)

\smallskip
\textbf{Composition (not requested in training), B-T2-comp.} \emph{The cell has $Q_\text{nom}$ = 20 Ah, $k$ = 1/6, and $I_\text{ref}$ = 4 A. At time 0, its SOC is 7/10. A constant discharge current $I$ = 9 A flows from time 0 to time 2100 s. Then a constant discharge current $I$ = 16 A flows from time 2100 s to time 3000 s. Calculate the SOC at the end of the second discharge.}
Key: each segment has its own usable charge (15 Ah, then 40/3 Ah); final SOC $= 7/10 - 7/20 - 3/10 = 1/20$. (Constant capacity: 0.2375; the first current's $Q_\text{use}$ for both segments: 0.083.)
\end{minipage}}
\caption{\textbf{The same rule, used once or twice.} Two fresh test items (abridged). The composition item needs T2 at two currents; no training question contains a second segment.}
\label{fig:items}
\end{figure}

\paragraph{Item families.}
\textbf{A} (forward use, one segment): compute $v$ and the terminal voltage (A-T1), or $Q_\text{use}$ and the final SOC (A-T2).
\textbf{B-inv} (inversion): recover $\tau$ from an observed voltage, or $k$ from an observed SOC drop.
\textbf{B-comp} (composition): the same rule over two consecutive segments, where the state at the end of the first segment starts the second: a current segment followed by a zero-current rest under the same relaxation law (B-T1-comp), or two discharge segments at different currents, each with its own $Q_\text{use}$ (B-T2-comp; Figure~\ref{fig:items}).
Composition is therefore the sequential reuse of one rule, not a combination of T1 with T2.
\textbf{C-abst} (abstention): a question about a layer whose law is not given; the correct answer says that the value is not determined.
The 128 training questions are A items, 64 per rule.
Throughout, \textbf{``unpracticed'' means not requested by the training questions}: no training question asks for an inversion, a rest segment or a second current.
We do not claim that no training \emph{target} contains such a step or one of its substeps; the targets were scanned with patterns, but a complete semantic reading was not done (Section~\ref{sec:limits}).

\paragraph{Item sets and endpoints.}
The \emph{sealed} set (124 items: 48 A, 24 B-inv, 24 B-comp, 12 C-abst and 16 recall and concept probes) was generated with the training items before any training.
The \emph{fresh} set (120 items: 24 A, 48 B-inv, 48 B-comp) was generated later from the same seeded stream for a prospective test; no fresh world repeats an earlier world of the same rule.
Sealed and fresh items use rule arguments that also occur in training questions: $h/\tau \in \{1/4, 1/3, 1/2, 1\}$ and $\sqrt{I/I_\text{ref}} \in \{1, 3/2, 2\}$.
The \emph{new-argument} set (120 items with the same family counts) was generated before any of its requests; every T1 segment ratio and every T2 current ratio in it is new, and 39 of its 60 T2 items also have a new product $k\sqrt{I/I_\text{ref}}$.
\textbf{The composition endpoint of each 120-item panel is its 48 B-comp items.}
An arm with three training seeds gives 144 composition answers and an arm with two seeds gives 96; these answers are repeated measurements of 48 items, not independent items.

\subsection{Capture policies, selection and training}

\paragraph{Capture.}
With Thinking enabled, the base model answers each training question twice under each capture context (Figure~\ref{fig:story}a).
In \genT{}, the system message contains the rule sheet; in \instT{}, it contains the same sheet and a program-written worked solution of that question, which states the factor, the intermediate values and the final answer.
An automated judge (Claude Opus 5.5 at high effort, in a separate context) accepts an answer if its final answer is correct against the program key; acceptance does not certify every intermediate statement.
The training target is the whole completion, including the Thinking text.
\genT{} has accepted first-sample targets for all 128 questions and \instT{} for 127.
For the replication (Section~\ref{sec:banks}), each policy answered every training question twice more with new request seeds, and the same judge and rule produced a second, independent pair of target banks.
An answer-only policy (\ansT; Section~\ref{sec:ans}) was captured in the same way, with the sheet and only the final-answer line of the worked solution.
A kept-sheet control (\keptT) trains on the \genT{} targets with the sheet left in the training input.

\paragraph{Training.}
Each pass presents one accepted target per question with the context removed; passes alternate between the two accepted samples.
The recipe is the same for every policy: LoRA+ \citep{hu2022lora,hayou2024loraplus} of rank 16 on all seven projection matrices of every layer (25.1M trainable parameters), learning rates $10^{-5}$ and $4\times10^{-5}$ for the two factors, warm-up, constant phase and cosine decay within each pass, token cross-entropy over the whole completion, and a KL penalty (weight 0.05) to the base model on replay answers.
An update holds at least 4,096 supervised domain tokens, and each pass continues from the adapter and optimizer state of the previous pass with a new learning-rate schedule.
Weights and the key--value cache are BF16, the adapter and optimizer are FP32, and nothing is quantized.

\paragraph{Models.}
MiniCPM5-2B (revision \texttt{12a3808a}) is the main model.
Qwen3-4B-Thinking-2507 (revision \texttt{768f209d}) repeats the \genT{} acquisition result only.
Before training, each model had to pass a gate: with the sheet in its context it must solve each item family, and without the sheet it must fail.
Qwen3-1.7B failed this gate (0 of 8 rest-segment items with the sheet) and was not trained.

\subsection{What the comparison identifies}
\label{sec:identifies}

The worked solution adds no rule specification beyond the sheet: a deterministic program computes it from the sheet and the question.
It does change the distribution of the generated targets and, with it, the training exposure: target length, content (including any incidental steps of unrequested operations), references to the solution and the number of supervised tokens per pass.
\textbf{The experiment measures the downstream contrast between two complete capture-and-training policies; it does not identify which of these changes mediates the contrast.}
Where we report equal forward rates, they are observed rates, not evidence of equal mastery.

\subsection{Measures and statistics}

Each trained state answers each item once without any context, with up to 32,768 new tokens and a fixed seed per item that is shared across states.
Sampling follows each model's published configuration with a repetition penalty of 1.05 (MiniCPM5-2B: temperature 1.0, top-$p$ 0.95); capture uses the same settings.

\paragraph{Value signature} (pre-specified primary measure).
A program searches the final answer for each keyed quantity; a number matches if it lies within a quarter of the generator's separation margin of the Kestrel value.
A quantity whose Kestrel value also appears among the numbers of the question is set aside as ambiguous.
An answer succeeds if at least one keyed quantity is unambiguous and every unambiguous quantity has its Kestrel value.
The signature checks numbers, not derivations, so a correct number in a wrong role can pass.

\paragraph{Official endpoint} (secondary layer).
The same judge model reads the final answer with the item key and a rubric; success requires a normal completion, an English final answer, a correct conclusion, correct supporting values and no material contradiction.
Each judge packet, read in its own context, holds the answers of several states to the same item under opaque identifiers, with four answers of known verdict as sentinels.
An answer without an assessment (for example, without a final answer) is a failure in every denominator.
The two measures read the same answers and agree on 92.5--98.6\% of assessed answers per population (Appendix~\ref{app:agreement}); this agreement is not an independent replication.
All judgments in this study are automated; no answer was graded by a human.

\paragraph{Statistics.}
For each item, we average success over the training seeds of an arm and compare arms by the paired per-item difference: its mean, a 95\% percentile interval from 10,000 item-bootstrap draws (seed 20261011), and an exact two-sided sign-flip test, with Holm correction within each pre-specified family.
A pre-specified test \emph{holds} if the difference reaches its recorded margin, the interval excludes zero and the (Holm-adjusted) $p$ is below 0.05.
The intervals describe variation over items given the trained states.
They do not cover variation between training runs or target banks, which we examine separately with the seed-level results (Appendix~\ref{app:variation}), nor variation between models or domains.
The sign-flip test assumes sign-symmetric item contrasts; policies were not randomized to items.
We label every result \emph{pre-specified} (test and margin fixed before the data), \emph{prospective confirmation} (a hypothesis from earlier data tested on new items with a fixed rule), \emph{sensitivity}, \emph{descriptive} or \emph{exploratory} (chosen after the data).
Appendix~\ref{app:prereg} lists every pre-specified test with its outcome and the dates on which it was recorded.

\section{Equal forward accuracy, unequal composition}
\label{sec:results}

\subsection{Self-generated targets put the new rules into the weights}
\label{sec:dose}

The base model never produces a Kestrel value without the sheet (0 of 48 sealed forward answers in each of two draws, and 0\% on the fresh and new-argument forward items).
With \genT{} targets, official forward success on the sealed items is 8.3\% after two passes, 63.5\% after four and 82.3\% after six (descriptive contrast recorded in the protocol: six minus two passes $+74.0$\pp{} \ci{63.9}{83.3}), and inversion and composition rise with it (Appendix~\ref{app:acquisition}, Table~\ref{tab:dose}).
With the sheet kept in the training input (\keptT), the same targets give 0\% forward success in each of two seeds, a local replication of contextual shadowing \citep{park2025newnews,yang2025rule}.
On Qwen3-4B-Thinking-2507, six \genT{} passes raise forward success from 16.7\% after two passes to 81.2\% ($+64.6$\pp{} \ci{53.1}{76.0}), but its transfer to unpracticed operations misses its pre-specified margin ($+19.8$\pp{} against 20), so this model supports the acquisition result only, and we did not run the policy comparison on it.
Use training also has side effects: it removed correct abstention on the 12 tested questions about a layer whose law is not given (base model 4 of 12 in each draw; six \genT{} passes 0 of 12 in each seed), and one six-pass seed reached the pre-specified damage threshold on a 56-item retention panel (Appendix~\ref{app:acquisition}).

\subsection{Forward accuracy does not reveal the capture context; composition does}
\label{sec:main}

\paragraph{Discovery.}
We trained \instT{} to test whether a worked solution in the capture context changes forward acquisition.
The pre-specified superiority test did not reject: after six passes both policies have 82.3\% official forward success on the sealed items ($0.0$\pp{} \ci{-11.5}{11.5}), an interval too wide to show equal acquisition.
On the same states, composition success differed by $+33.3$\pp{} \ci{16.7}{50.0}.
Because we found this difference after seeing the data, we treated it as a hypothesis and tested it prospectively.

\paragraph{Prospective confirmation.}
Before any new request, we recorded two tests on 120 newly generated items and a third training seed of each policy (Appendix~\ref{app:prereg}): TB-1 (six \genT{} passes minus six \instT{} passes) and TB-2 (six \genT{} passes minus ten \instT{} passes; Section~\ref{sec:dosebundle}), both on the 48 fresh composition items, with the value signature, a 10-point margin and Holm correction over the two.
TB-1 holds: $+24.3$\pp{} \ci{14.6}{34.0}, Holm $p < 0.001$; the sheet-only policy is ahead on 29 items and behind on 5 (Table~\ref{tab:fresh}).
The official judge, applied to all 288 composition answers, agrees in direction: 82 of 144 \genT{} answers succeed against 45 of 144 \instT{} answers, $+25.7$\pp{} \ci{16.0}{35.4}.
The lead is positive in each of the three seed pairs ($+14.6$, $+33.3$ and $+25.0$; Appendix~\ref{app:variation}) and on both rules; it is largest on the two-current items ($+30.6$\pp{} \ci{20.8}{40.3}) and smallest on the rest-segment items ($+18.1$\pp{} \ci{1.4}{34.7}).
Inversion, the other unpracticed operation, shows a lead of $+16.7$\pp{} \ci{4.9}{28.5}, while the fresh forward rates are 86.1\% and 83.3\% ($+2.8$\pp{} \ci{-9.7}{15.3}).
An exploratory contrast of contrasts (composition minus forward difference, per item) gives $+21.5$\pp{} \ci{5.6}{36.8}; it does not control latent mastery, and the item sets were not randomized to operations.

\begin{table}[tbp]
\centering
\caption{\textbf{The composition gap holds prospectively, and ten passes of the worked-solution policy do not close it.} Fresh items, bank pair~1 (MiniCPM5-2B, no context). Rates are per-item means over training seeds (number of seeds in brackets); differences are \genT{} minus \instT{} in percentage points, with 95\% item-bootstrap intervals. TB-1 and TB-2 are pre-specified (value signature, 10-point margin, Holm over the two); TB-2 is a sensitivity comparison (Section~\ref{sec:dosebundle}). Official: semantic judge.}
\label{tab:fresh}
\footnotesize
\setlength{\tabcolsep}{3.5pt}
\begin{tabular}{llccccc}
\toprule
Items & Measure & \genT{} 6 [3] & \instT{} 6 [3] & \instT{} 10 [2] & \genT{} 6 $-$ \instT{} 6 & \genT{} 6 $-$ \instT{} 10 \\
\midrule
Composition (48) & signature & 59.0 & 34.7 & 42.7 & TB-1: $+24.3$ \ci{14.6}{34.0} & TB-2: $+16.3$ \ci{6.3}{26.7} \\
Composition (48) & official & 56.9 & 31.2 & 38.5 & $+25.7$ \ci{16.0}{35.4} & $+18.4$ \ci{8.0}{28.8} \\
Inversion (48) & signature & 66.7 & 50.0 & 54.2 & $+16.7$ \ci{4.9}{28.5} & $+12.5$ \ci{0.3}{25.3} \\
Forward (24) & signature & 86.1 & 83.3 & 85.4 & $+2.8$ \ci{-9.7}{15.3} & $+0.7$ \ci{-15.3}{17.4} \\
\bottomrule
\end{tabular}
\end{table}

\subsection{Not only less training}
\label{sec:dosebundle}

The most direct objection is dose, and this section asks how much of the gap it explains.
Worked-solution targets are shorter (the median capture response has 1,868 generated tokens with the worked solution and 3,105 without), so six \instT{} passes supervise 44\% fewer domain tokens than six \genT{} passes (Appendix~\ref{app:acquisition}, Table~\ref{tab:bundle}).
Ten \instT{} passes come within 6.6\% of the domain tokens of six \genT{} passes and reach nearly the same fresh forward rate (85.4\% against 86.1\%).
The composition lead persists: TB-2 holds ($+16.3$\pp{} \ci{6.3}{26.7}, Holm $p = 0.005$; official $+18.4$\pp{} \ci{8.0}{28.8}, 82 of 144 against 37 of 96 answers), and on the two training seeds that both arms share the lead is $+16.7$\pp{} \ci{4.2}{29.2}.
Dose matters: four more passes raise \instT{} composition from 34.7\% to 42.7\%.
Within the range we tested, it does not close the gap.
Ten passes also add replay tokens, target visits, updates and learning-rate cycles, so \textbf{TB-2 is a sensitivity comparison: the lead persists at about matched domain tokens and observed forward rate, but no single dose quantity is isolated.}
The same picture holds along the sealed training ladders (Figure~\ref{fig:curves}, left): the \instT{} states lie below the \genT{} curve at every forward rate that both reach, and ten passes move \instT{} up its own curve rather than onto the \genT{} curve; over forward rates of 50--80\%, the mean vertical distance between the curves is $+18.7$\pp{} \ci{2.4}{34.9} (exploratory; the statistic was defined after the answers existed).
Section~\ref{sec:account} explains why this pattern can arise and what it predicts; Section~\ref{sec:banks} tests the dose result with a second bank pair.

\section{Why equal practice accuracy can hide different transfer}
\label{sec:account}

We wrote this account after the results of Section~\ref{sec:results} were known, so its agreement with them is retrospective.
Its predictions for the tests of Section~\ref{sec:tests}, and those of rival accounts, were recorded before any answer of those tests existed (Appendix~\ref{app:prereg}).
It is an idealized surrogate: it shows that one simple mechanism produces the pattern, not that the network uses it.

\subsection{Two routes, one direction}

Idealize the trained model's answer to a forward question as a linear readout of two routes to the requested value.
A \emph{general route} $g(x)$ evaluates the rule from the arguments and can be reused when the configuration changes.
A \emph{support-bound route} $h(x)$ gives the correct value only on practice-like questions, for example a forward procedure that works only in the trained configuration.
On every training question both routes give the target value, scaled to $g = h = 1$, and the output is $f = u\,g + v\,h$.
The targets supervise the two routes with gains $\alpha$ and $\beta$, so that $u = \alpha\theta_1$ and $v = \beta\theta_2$ for trained parameters $\theta$ whose effective update starts at zero, as a LoRA update does.
With a squared surrogate loss on the practiced output and the replay penalty idealized as an isotropic quadratic penalty with weight $\lambda > 0$,
\begin{equation}
L(\theta) = \tfrac12\,(\alpha \theta_1 + \beta \theta_2 - 1)^2 + \tfrac{\lambda}{2}\,\lVert\theta\rVert^2 .
\label{eq:loss}
\end{equation}

\begin{proposition}[the targets fix the direction]
\label{prop:direction}
Let $a = (\alpha, \beta)$ with $\rho^2 = \alpha^2 + \beta^2 > 0$, and let $\theta$ follow the gradient flow $\dot\theta = -\nabla L(\theta)$ from $\theta(0) = 0$.
Then $\theta(t) = \theta^*\,(1 - e^{-(\rho^2+\lambda)t})$ with $\theta^* = a/(\rho^2+\lambda)$, so $\alpha^2 v(t) = \beta^2 u(t)$ for all $t$, and $v(t)/u(t) = \beta^2/\alpha^2$ for all $t > 0$ when $\alpha \neq 0$.
The practiced output is $m(t) = u(t) + v(t)$, and where $h$ does not apply the output is $c\,u(t) = k\,m(t)$, with $k = c\,\alpha^2/\rho^2$ and an efficiency $c \le 1$ of the general route in that use.
\end{proposition}

\begin{proof}
$\nabla L(\theta) = a\,(a^\top\theta - 1) + \lambda\theta$, so the Hessian $a a^\top + \lambda I$ has eigenvalue $\rho^2 + \lambda$ along $a$ and $\lambda$ orthogonal to it, and the minimizer is $\theta^* = a/(\rho^2+\lambda)$.
The deviation $\theta(t) - \theta^*$ starts at $-\theta^*$, which lies along $a$, and decays as $e^{-(\rho^2+\lambda)t}$ along that eigenvector, which gives $\theta(t)$.
Then $u = \alpha\theta_1 = \alpha^2 s(t)$ and $v = \beta\theta_2 = \beta^2 s(t)$ with $s(t) = (1 - e^{-(\rho^2+\lambda)t})/(\rho^2+\lambda)$, so $m = \rho^2 s(t)$ and $c\,u = c\,(\alpha^2/\rho^2)\,m$.
\end{proof}

\noindent
\textbf{Training follows a direction that the targets set, and more training only moves along it.}
Two policies can therefore reach the same practiced output with different splits between the routes, and different outputs where the support-bound route does not apply.
The surrogate idealizes a great deal: the LoRA update is bilinear in its factors, not linear in $\theta$, and LoRA+ learning-rate asymmetry, adaptive optimization, token-level cross-entropy and a KL replay penalty do not reduce to this flow; the routes are not identified in the network.

\subsection{From outputs to accuracy: orderings, not ratios}

The surrogate describes continuous outputs, while we measure success rates.
Reading a ratio of success rates as an estimate of $k$ would be invalid, because success is a threshold or link function of the output: with practiced output 0.9, transfer output 0.45 and a success threshold of 0.4, the output ratio is 0.5 and the success ratio is 1.
A monotone observation model connects the two without estimating anything.
Let forward accuracy be $F = \Phi_F(m)$ and composition accuracy $C = \Phi_C(k\,m)$.
We assume policy-invariant, strictly increasing observation links $\Phi_F$ and $\Phi_C$ over the range considered.
This is an additional modeling assumption, not a consequence of using the same evaluation protocol: the items, decoding and scoring are shared, but a policy can change answer format or error types, and with them the links.

\begin{proposition}[orderings]
\label{prop:order}
Under this observation model, the states of a policy lie on the curve
\begin{equation}
C = h_k(F) = \Phi_C\!\left(k\,\Phi_F^{-1}(F)\right),
\label{eq:curve}
\end{equation}
which is strictly increasing in $k$ at every $F$ reached with $m > 0$.
\end{proposition}

\noindent
The proof is immediate: $\Phi_F^{-1}$ exists on the forward accuracies reached because $\Phi_F$ is strictly increasing, $m = \Phi_F^{-1}(F) > 0$, and $\Phi_C$ is strictly increasing.
Three predictions follow, none of which needs the form of the links:
\begin{itemize}
\item[R1] \emph{Ordered curves.} Each policy traces one curve of composition against forward accuracy, and curves with larger $k$ lie above at every forward accuracy that both reach; they do not cross.
\item[R2] \emph{Dose moves along the curve.} More passes on the same targets move a state along its own curve, so a policy with a lower curve stays lower at matched forward accuracy, whatever its dose.
\item[R3] \emph{Speed alone collapses.} Policies that differ only in learning speed ($\rho$, at fixed $\alpha/\rho$) share $k$ and lie on one curve.
\end{itemize}
R2 and R3 turn the dose objection into a test.
If the worked solution only slowed learning, a worked-solution state that reaches the forward accuracy of a sheet-only state would also reach its composition accuracy.
TB-2 compared such states (86.1\% and 85.4\% forward success) and found a composition lead of 16.3 points, as R2 requires and R3 does not; because TB-2 was recorded before the account existed, this agreement is retrospective.
The common-link assumption could fail if one policy changed answer format or effort; the worked-solution models write longer composition answers, not shorter ones (Section~\ref{sec:diagnostics}).

\subsection{What the support-bound route is bound to}
\label{sec:binding}

The surrogate does not say what $h$ is bound to, and the forms that it leaves open predict different results at argument values that no training question used.
If $h$ holds values or templates of the trained arguments (\emph{argument-bound}), forward use fails at new arguments in the policy that relies more on $h$, so a forward gap appears.
If $h$ is a forward procedure that works at any arguments but only in the trained configuration (\emph{configuration-bound}), forward use survives at new arguments in both policies while the composition gap persists.
In a \emph{value-association} alternative, neither policy holds a rule that works at new values: both fall to or near the floor at new arguments, and within T2, items whose product $k\sqrt{I/I_\text{ref}}$ occurs in training do better than items with a new product.
We recorded these predictions with probabilities before any answer of the new-argument test existed (Table~\ref{tab:predictions}): a forward gap of at least 10 points (argument-bound, 0.35); a forward gap within $\pm10$ points (configuration-bound, 0.55); a composition gap larger than the forward gap (0.60); and trained products at least 15 points easier than new ones (value association, 0.35).
The prediction was not obvious: worked-solution targets assert values at the training arguments, so an argument-bound shortfall was a live possibility.

Later we recorded predictions for a second bank pair (the direction of the gap should recur), for that pair at matched dose (R2: the gap should persist, recorded with probability 0.30 that the test would reach its margin), and for an answer-only context.
For the answer-only context, the account as we then stated it located the cost in values that targets confirm against supplied intermediate results, so its most probable recorded reading (0.40) was a position near the sheet-only policy: without a supplied method, the targets would have to compute the rule themselves.
An \emph{absent-reference} account predicted the opposite, a position near the worked-solution policy, because targets that rely on a supplied result teach the model to rely on a result that is absent later.
The strongest alternative that none of these tests can exclude is a package: the worked-solution targets are shorter, deliberate and check less, cite an absent solution and supply fewer tokens per pass, and any combination of these could produce the gap.

\section{Three tests recorded before their data}
\label{sec:tests}

\subsection{New argument values preserve forward accuracy and the composition gap}
\label{sec:offsup}

Every sealed and fresh item reuses rule arguments of the training questions, so a model could answer with values or templates bound to those arguments, and the two policies could differ in how much they do so.
This test asks whether the gap is a difference in recall of the trained argument values.
Before any request, we generated the new-argument set (Section~\ref{sec:setting}) and recorded two tests with their reading rules and the predictions of Section~\ref{sec:binding}: OS-1, six \genT{} passes minus the base model on the 24 forward items, with a \textbf{30-point} margin; and OS-2, six \genT{} passes minus six \instT{} passes on the 48 composition items, with a 10-point margin (value signature, Holm over the two).
The six-pass states of the original banks, three seeds per policy, answered without any context.

Both tests hold (Table~\ref{tab:offsup}).
Forward success is 87.5\% for \genT{} against 0\% for the base model (OS-1: $+87.5$\pp{} \ci{80.6}{93.1}; official \genT{} 86.1\% and base model 0\%, the only two states that the official forward layer covers).
\textbf{The observed forward rates of the two policies are identical, 87.5\% each ($0.0$\pp{} \ci{-9.7}{9.7}); by rule they differ by $+8.3$ (T1) and $-8.3$ points (T2), so these comparisons do not establish equivalent mastery.}
On composition, sheet-only targets lead by 34.7 points (OS-2: \ci{25.7}{43.8}, Holm $p < 0.001$; official $+34.7$\pp{} \ci{25.7}{43.8}), in each of the three seed pairs ($+29.2$, $+35.4$ and $+39.6$) and on both rules ($+38.9$ on rest-segment and $+30.6$ on two-current items).
With the sheet in its context, the base model answers 98.3\% of these items (fresh items: 92.5\%): the new-argument items remain highly solvable when the rules are supplied.
Within T2, \genT{} answers are not more successful on items whose product $k\sqrt{I/I_\text{ref}}$ occurs in training (55.6\% of 21 items) than on items with a new product (59.8\% of 39).
On the second model, six \genT{} passes also answer 87.5\% of the new-argument forward items (base model 0\%; with the sheet, 100\%).

The configuration-bound form predicted this pattern; the argument-bound form (a forward gap of at least 10 points) and the value-association form (trained products at least 15 points easier) did not.
\textbf{This is the account's one prospective confirmation that discriminates between explanations.}
It rules against pure lookup of the original argument grid as the source of the gap; it does not exclude mixed template-and-value mechanisms, and it does not show general symbolic extrapolation.
The gap is larger here than on fresh items because the worked-solution policy is lower on these items (23.6\% against 34.7\%), while the sheet-only policy is not (58.3\% against 59.0\%); the item sets differ, so this comparison is descriptive.

\begin{table}[tbp]
\centering
\caption{\textbf{At argument values that no training question used, forward use survives in both policies and the composition gap remains.} New-argument items (MiniCPM5-2B, no context; three seeds per policy, per-item means). OS-1 (30-point margin) and OS-2 (10-point margin) are pre-specified. The official layer covers the forward answers of \genT{} and the base model and all composition answers.}
\label{tab:offsup}
\footnotesize
\setlength{\tabcolsep}{4pt}
\begin{tabular}{llcccc}
\toprule
Items & Measure & \genT{} 6 & \instT{} 6 & Base model & \genT{} 6 $-$ \instT{} 6 \\
\midrule
Forward (24) & signature & 87.5 & 87.5 & 0.0 & $0.0$ \ci{-9.7}{9.7} \\
Forward (24) & official & 86.1 & -- & 0.0 & -- \\
Inversion (48) & signature & 62.5 & 47.9 & 2.1 & $+14.6$ \ci{2.8}{26.4} \\
Composition (48) & signature & 58.3 & 23.6 & 2.1 & OS-2: $+34.7$ \ci{25.7}{43.8} \\
Composition (48) & official & 54.9 & 20.1 & -- & $+34.7$ \ci{25.7}{43.8} \\
\bottomrule
\end{tabular}
\end{table}

\subsection{A second bank pair: the direction recurs; the matched-dose test fails}
\label{sec:banks}

Training seeds vary the order and sampling of training, not the capture, so the gap could belong to one pair of sampled banks.
We recorded this replication before any of its requests (amendments 7 and~8): we captured two new samples per question for each policy, with the same questions, contexts and settings and new request seeds, accepted and selected them by the same rules, and trained six passes with the first two training seeds of the original arms.
The test RB-1 (value signature, 10-point margin) compares the new banks on the 48 fresh composition items.

RB-1 holds (Table~\ref{tab:banks}): sheet-only targets lead by 14.6 points \ci{2.1}{26.0} ($p = 0.033$; official $+16.7$\pp{} \ci{5.2}{28.1}, 42 of 96 against 26 of 96 answers), and, as recorded in advance, the lead is larger on two-current ($+22.9$) than on rest-segment items ($+6.2$).
\textbf{The observed forward rates differ by 8.3 points (87.5\% against 79.2\%, \ci{-10.4}{27.1}); the recorded $\pm10$-point gate is a point-estimate rule, not an equivalence test, so at six passes this replication comes with a forward shortfall of the worked-solution bank.}
To remove that shortfall, we recorded a dose extension before any of its requests: the new worked-solution bank trained to ten passes (97\% of the domain tokens of the new sheet-only bank), with the TB-2 rule (test TB-2R).
Forward success was then 89.6\% against 87.5\%, and composition 39.6\% against 45.8\%: $+6.2$ points \ci{-6.3}{18.8} ($p = 0.43$; official $+7.3$\pp{} \ci{-5.2}{19.8}).
\textbf{TB-2R does not hold.} For this bank pair, R2 is not supported at the recorded margin, and a difference in learning speed alone (R3) is not excluded.

Does the second bank pair carry a smaller cost, or is it the same cost measured with noise?
Our data cannot decide (Appendix~\ref{app:variation}; exploratory).
The gaps of the two bank pairs differ by $+9.7$\pp{} \ci{-6.6}{27.1} at six passes and $+10.1$\pp{} \ci{-5.6}{25.7} at matched dose, and seed pairs within one bank pair differ by as much: from $+14.6$ to $+33.3$ points within bank pair~1 at six passes, and $+16.7$ and $-4.2$ points within bank pair~2 at matched dose.
The two sheet-only banks reach 59.0\% and 45.8\% composition success, but trained with the same seed they differ by $+2.1$ points \ci{-14.6}{18.8} for one seed and $+25.0$ points \ci{6.3}{43.8} for the other.
With two banks per policy and two or three seeds per bank, variation between banks cannot be separated from variation between training runs.
Pooled over both bank pairs, with each pair weighted equally per item (a descriptive estimate named in advance; the equal weighting was fixed when it was first computed, after the six-pass results), the lead is 19.4 points \ci{12.3}{26.2} at six passes and 11.3 points \ci{2.4}{19.8} at matched dose (official $+21.2$ and $+12.8$); these intervals are conditional on the four sampled banks and contain neither bank nor training-run variation.
\textbf{The direction of the gap recurs in both bank pairs and at both doses; at matched dose its size is uncertain, 6.2 and 16.3 points for the two bank pairs.}

\begin{table}[tbp]
\centering
\caption{\textbf{An independently sampled bank pair repeats the direction of the gap at a smaller size; at matched dose its own gap is not distinguishable from zero.} Fresh items, value signature unless stated; per-item means over training seeds (number of seeds in brackets). Bank~1 is the bank of Table~\ref{tab:fresh}. Lower block: the worked-solution policy of each bank trained for ten passes (about the domain tokens of six sheet-only passes); the bank-2 composition row is the test TB-2R. Pooled: mean of the two bank-pair differences per item, with a 95\% item-bootstrap interval (descriptive, conditional on the four banks).}
\label{tab:banks}
\footnotesize
\setlength{\tabcolsep}{3.2pt}
\begin{tabular}{lccccccc}
\toprule
 & \multicolumn{2}{c}{Bank 1} & \multicolumn{2}{c}{Bank 2} & \multicolumn{3}{c}{\genT{} $-$ \instT{}} \\
\cmidrule(lr){2-3}\cmidrule(lr){4-5}\cmidrule(lr){6-8}
Items & \genT{} [3] & \instT{} [3] & \genT{} [2] & \instT{} [2] & Bank 1 & Bank 2 & Pooled \\
\midrule
Forward (24) & 86.1 & 83.3 & 87.5 & 79.2 & $+2.8$ & $+8.3$ & $+5.6$ \ci{-5.6}{17.4} \\
Inversion (48) & 66.7 & 50.0 & 60.4 & 33.3 & $+16.7$ & $+27.1$ & $+21.9$ \ci{13.9}{30.0} \\
Composition (48) & 59.0 & 34.7 & 45.8 & 31.2 & $+24.3$ & $+14.6$ & $+19.4$ \ci{12.3}{26.2} \\
Composition, official & 56.9 & 31.2 & 43.8 & 27.1 & $+25.7$ & $+16.7$ & $+21.2$ \ci{14.6}{27.6} \\
\midrule
\multicolumn{8}{l}{\emph{Matched dose: worked-solution policy trained for ten passes} [2]} \\
Forward (24) & 86.1 & 85.4 & 87.5 & 89.6 & $+0.7$ & $-2.1$ & $-0.7$ \ci{-10.8}{10.8} \\
Inversion (48) & 66.7 & 54.2 & 60.4 & 58.3 & $+12.5$ & $+2.1$ & $+7.3$ \ci{-0.3}{15.1} \\
Composition (48) & 59.0 & 42.7 & 45.8 & 39.6 & $+16.3$ & $+6.2$ & $+11.3$ \ci{2.4}{19.8} \\
Composition, official & 56.9 & 38.5 & 43.8 & 36.5 & $+18.4$ & $+7.3$ & $+12.8$ \ci{4.2}{21.2} \\
\bottomrule
\end{tabular}
\end{table}

\begin{figure}[tbp]
\centering
\includegraphics[width=\linewidth]{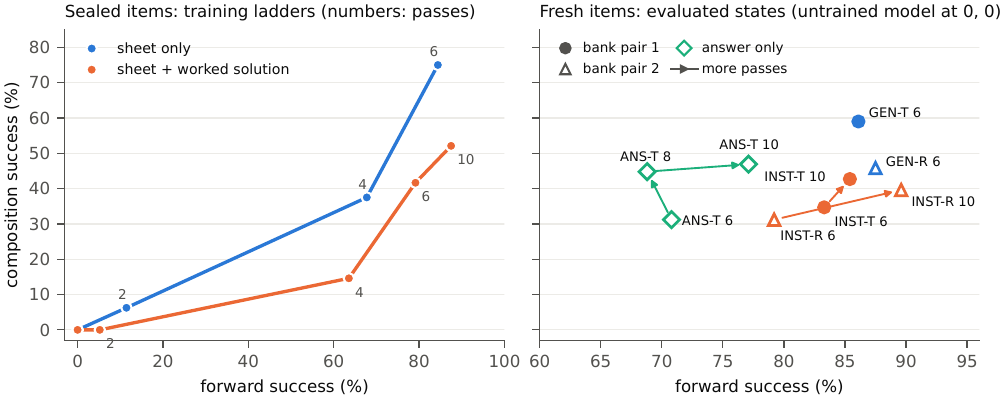}
\caption{\textbf{More training moves the worked-solution policy along its own curve; with bank pair~1 that curve stays below the sheet-only policy, and with bank pair~2 it comes close.} Composition against forward success (value signature). Left: sealed items along the training ladders of bank pair~1 (means of training seeds 1 and~2; the untrained model is the first point; numbers are passes); the mean vertical distance between the curves over forward rates of 50--80\% is exploratory. Right: fresh items at every evaluated state; arrows join states of one policy and bank after more passes (worked-solution policy: six to ten passes; answer-only policy: six, eight and ten passes).}
\label{fig:curves}
\end{figure}

\subsection{An answer-only context: the mediator stays open}
\label{sec:ans}

Does the cost need the supplied method, or does a supplied final answer suffice?
A third policy (\ansT) captured targets with the rule sheet plus the line ``Checked final answer of this question:'' and the final-answer line of the worked solution, with no method and no intermediate value; capture, acceptance, selection and training followed the second bank pair (two seeds).
Before any answer, we recorded a reading table and a gate: the comparisons AS-1 (\genT{} 6 minus \ansT{} 6) and AS-2 (\ansT{} 6 minus \instT{} 6), on the original banks' states, would be read only if \ansT{} forward success was within 15 points of \genT{}.
\textbf{It was not: the answer-only policy reached 70.8\% against 86.1\%, 15.3 points below, so no reading applies.}
Descriptively, its composition success was 31.2\% (\instT{} 34.7\%, \genT{} 59.0\%; AS-1 $+27.8$\pp{} \ci{17.7}{37.8}, AS-2 $-3.5$\pp{} \ci{-14.9}{8.3}).
A lower forward rate puts a state lower on its policy's curve, so the failed gate made the result uninformative about the account's prediction in either direction.

We then recorded, before any request, a dose extension of \ansT{} to eight passes (103\% of the domain tokens of six \genT{} passes), read with the unchanged table, and a fixed second look at ten passes (129\%) if eight passes failed the gate.
\textbf{Eight passes failed the gate again}: forward success 68.8\%, 17.4 points below, while composition rose to 44.8\%.
Ten passes passed it (77.1\%, 9.0 points below) and gave the ``graded'' row of the table: \ansT{} 10 composition (46.9\%) is 12.2 points \ci{0.7}{23.6} below \genT{} 6 and 12.2 points \ci{0.7}{23.6} above \instT{} 6.
The support for this reading is weak.
Both lower bounds are 0.7 points after two looks without adjustment.
The official judge places \ansT{} 10 nearer \instT{} 6 ($+18.4$\pp{} \ci{7.3}{29.5} and $+7.3$\pp{} \ci{-4.9}{19.8}), the pattern of the row in which the answer alone gives most of the cost.
The recorded comparators are states of the original banks, and against the second bank pair, captured in the same campaign as \ansT{}, its composition rate is within 7.3 points of both policies (exploratory, chosen after the result).
From six to eight passes, \ansT{} composition rose by 13.5 points \ci{-1.0}{27.1} while its forward rate did not rise (70.8\% and 68.8\%); movement along one curve does not produce that, so either the common link fails for this policy or the change is noise.
\textbf{The account's most probable reading, a position near the sheet-only policy, did not occur, and the data do not show whether a supplied final answer alone produces most of the cost.}

\subsection{Scorecard}

Table~\ref{tab:predictions} lists every recorded prediction with its probability and outcome.
The binding test is the account's one discriminating prospective confirmation.
The bank replication held as recorded at six passes, but the dose ordering (R2) held only for bank pair~1, where it was retrospective, and failed its margin for bank pair~2.
The predictions about the likelihood diagnostic $D$ held, but their sign is close to guaranteed by construction (Section~\ref{sec:diagnostics}), so they are a weak check.
The answer-only prediction did not hold.
These outcomes are correlated readings of a few experiments, not a combined statistical validation of the account.

\begin{table}[tbp]
\centering
\caption{\textbf{The binding prediction held; the dose-matched second bank pair and the answer-only policy did not follow the account.} Predictions recorded before each test, with their probabilities and outcomes (value signature; ``gap'': \genT{} minus \instT{} after six passes unless stated). Answer-only readings were recorded with probabilities 0.40 (near \genT{}: the cost needs the method), 0.20 (graded), 0.15 (near \instT{}: the answer alone gives the cost) and 0.25 (other or not comparable) before the six-pass test, and 0.25, 0.05, 0.35 and 0.35 before the dose extension.}
\label{tab:predictions}
\footnotesize
\setlength{\tabcolsep}{3pt}
\begin{tabular}{p{0.15\linewidth}p{0.43\linewidth}cp{0.27\linewidth}}
\toprule
Test & Prediction & Prob. & Outcome \\
\midrule
New arguments & forward gap at least $+10$\pp{} (argument-bound) & 0.35 & no: $0.0$ \ci{-9.7}{9.7} \\
 & forward gap within $\pm10$\pp{} (configuration-bound) & 0.55 & yes (observed rates; not equivalence) \\
 & composition gap minus forward gap positive & 0.60 & yes: $+34.7$ \\
 & T2: trained product at least $+15$\pp{} above new product (value association) & 0.35 & no: $-4.2$ \\
Likelihood $D$ & larger for \instT{} than for \genT{} targets, each rule & 0.90 & yes; sign close to guaranteed by construction \\
 & arm difference larger for T2 than for T1 & 0.65 & yes (two rules; weak) \\
Second banks & RB-1 holds & 0.55 & yes: $+14.6$ \ci{2.1}{26.0} \\
 & forward gap within $\pm10$\pp{} & 0.70 & yes: $+8.3$ (point-estimate gate) \\
 & composition gap larger on T2 than on T1 & 0.65 & yes: $+22.9$, $+6.2$ \\
 & pooled composition gap at least $+15$\pp{} & 0.60 & yes: $+19.4$ \ci{12.3}{26.2} \\
Answer only & forward success within 15\pp{} of \genT{} & 0.75 & no: 15.3 below \\
 & reading of AS-1, AS-2 (see caption) & & not comparable \\
 & deliberation markers between the two policies & 0.70 & yes \\
Matched dose & second-bank \instT{} 10 forward within $\pm10$\pp{} of \genT{} 6 & 0.80 & yes: $-2.1$ \\
 & TB-2R holds (gap at least $+10$\pp{}, interval above 0) & 0.30 & no: $+6.2$ \ci{-6.3}{18.8} \\
 & TB-2R gap positive & 0.85 & yes \\
 & answer-only, eight passes, forward within 15\pp{} & 0.80 & no: 17.4 below \\
 & answer-only reading (see caption) & & graded, at ten passes \\
 & answer-only composition rises from six to eight passes & 0.80 & yes: $+13.5$ \\
\bottomrule
\end{tabular}
\end{table}

\section{What the targets and the trained models write}
\label{sec:diagnostics}

The analyses in this section were defined after the results were known.
They are program counts and readings of saved text, not tests, and they describe text, not internal mechanisms; they indicate which properties of the targets an edit experiment would have to separate.

\paragraph{Both policies write and evaluate the rules; they differ in what surrounds the computation.}
Table~\ref{tab:targets} describes every accepted training target of the original banks.
Both policies write each rule symbolically in every target and evaluate it with the question's numbers in 84--96\% of targets, so the worked solution does not remove the computation from the targets.
\instT{} targets name the worked solution in 81--96\% of cases, state in 42--47\% that a computed value matches it, and contain fewer markers of the model's own deliberation and checking; the reduction is strongest for T2, the rule with the larger composition gap.
The second bank pair repeats the pattern (median deliberation markers 18 and 24 in the sheet-only targets and 9 and 4 in the worked-solution targets, for T1 and T2).
In \genT{} targets, marker count and length are almost collinear ($r = 0.93$ and $0.96$), so these data cannot separate deliberation from length, and none of these counts is shown to mediate the gap.

\begin{table}[tbp]
\centering
\caption{\textbf{Both policies compute with the rules; the worked-solution targets deliberate and check less and confirm against a solution that will be absent.} Accepted training targets of bank pair~1, both samples (program counts defined after all results). ``Deliberation markers'': words such as wait, hmm and maybe; ``self-check phrases'': verify, double-check, recompute and similar. $D$: mean of $\log p_0(y \mid x, R, s) - \log p_0(y \mid x, R)$ per token under the base model $p_0$, one target per question, where $x$ is the question, $R$ the sheet and $s$ the worked solution.}
\label{tab:targets}
\small
\begin{tabular}{lrrrr}
\toprule
 & \genT{} T1 & \instT{} T1 & \genT{} T2 & \instT{} T2 \\
\midrule
Targets & 121 & 119 & 128 & 127 \\
Rule written symbolically (\%) & 100 & 100 & 100 & 100 \\
Rule evaluated with the question's numbers (\%) & 91.7 & 95.8 & 84.4 & 85.8 \\
Worked solution named (\%) & 0 & 80.7 & 0 & 96.1 \\
A value stated to match the solution (\%) & 0 & 42.0 & 0 & 46.5 \\
Median length (tokens) & 2,746 & 2,268 & 3,339 & 1,501 \\
Median deliberation markers & 18 & 10 & 26 & 4 \\
Median self-check phrases & 1 & 0 & 2 & 0 \\
$D$ (nats per token) & $-0.091$ & $+0.139$ & $-0.090$ & $+0.198$ \\
\bottomrule
\end{tabular}
\end{table}

\paragraph{The likelihood diagnostic $D$ records where a target came from, not what the student learned.}
Adding the worked solution to the sheet raises the base model's log-likelihood of every \instT{} target and lowers that of every \genT{} target (Table~\ref{tab:targets}), as we had predicted with probability 0.90.
This sign pattern is close to guaranteed by construction.
Let $q_s$ and $q_R$ be the base model's target distributions with and without the solution, and $D(y) = \log q_s(y)/q_R(y)$.
Then $\mathbb{E}_{q_s}[D] = \mathrm{KL}(q_s \,\|\, q_R) \ge 0$ and $\mathbb{E}_{q_R}[D] = -\mathrm{KL}(q_R \,\|\, q_s) \le 0$, so targets sampled with the solution tend to have positive $D$ and targets sampled without it negative $D$; sampling temperature, acceptance filtering and per-token normalization keep this from being exact.
$D$ therefore measures how far the solution moves the base model's distribution of targets.
It does not measure the supervision that reaches any route of the trained student, and because the solution is a deterministic function of the sheet and the question, it is not an estimate of a conditional mutual information.
\citet[Sec.~4.3]{tian2026selfteacher} likewise find that teacher--student discrepancy does not order privileged contexts by utility.

\paragraph{The worked-solution models expend more test-time tokens on composition, yet still fail more often.}
A natural suspicion is that training on shorter, solution-guided targets teaches the student model to generate shorter responses at test time.
On fresh compositional items, the opposite occurs: the median six-pass \instT{} response spans 8,837 generated tokens compared to 4,744 for \genT{}, whereas on practiced forward items the ordering reverses (2,748 versus 3,588 tokens).
The second bank pair exhibits the same pattern on composition (8,700 versus 5,349 median tokens).
\citet[Sec.~4.1]{lee2026answerconditioned} report a similar inversion of training and test lengths for answer-conditioned chains in mathematics; here the longer answers appear on the unrequested operation, not on the practiced one.

\paragraph{Under composition, both policies often fall back to the familiar law.}
On fresh two-current items, about two fifths of the answers of each six-pass policy carry the SOC that a constant capacity would give (40\% and 42\%), although both apply the rate-dependent usable charge in 89\% of six-pass forward T2 answers; the policies differ mainly in answers that match no known value (18\% against 43\%; Appendix~\ref{app:diagnostics}).

\paragraph{Worked-solution training leaves references to the absent solution.}
After training, \instT{} models refer to a worked solution or a sheet in the Thinking of 76--85\% of their answers, although none is present (six-pass \genT{}: 1--6\%; base model 2\%).
In joint teacher--student training, references to privileged information in student outputs did not grow with training \citep[App.~D.1]{penaloza2026pidistill}; offline training on targets that cite the solution does pass the citations on.
This is a pattern count; it does not show that the references cause the composition gap.

\paragraph{The rules appear in solutions, not on request.}
After six passes, both policies write the T1 factor in 86--100\% of fresh forward T1 answers, but none of nine answers of the six-pass \genT{} seeds to direct recall probes states either relation (Appendix~\ref{app:diagnostics}); we read this as elicitation behavior \citep{allenzhu2023physics31}, not as absent knowledge.

\section{What the choice costs in practice}
\label{sec:costs}

Capture and training costs describe only part of the policy trade-off; we also measure inference on unrequested composition.
Table~\ref{tab:costs} gives all three for each policy and bank pair, measured from the saved run records of one A40 GPU with one container image.

\begin{table}[tbp]
\centering
\caption{\textbf{The worked solution saves at capture and training and spends at inference.} Measured from the saved run records. Capture: base model, two samples per training question. Update time: sum of the measured optimizer-step times of all passes, mean over two seeds (model loading excluded). Inference: the fresh composition items, no context; ``tokens per success'' divides all generated composition tokens by the value-signature successes. The ten-pass states use the targets of the six-pass state of the same bank.}
\label{tab:costs}
\footnotesize
\setlength{\tabcolsep}{3pt}
\begin{tabular}{lrrrrrr}
\toprule
 & \multicolumn{3}{c}{Bank pair 1} & \multicolumn{3}{c}{Bank pair 2} \\
\cmidrule(lr){2-4}\cmidrule(lr){5-7}
 & \genT{} 6 & \instT{} 6 & \instT{} 10 & \genT{} 6 & \instT{} 6 & \instT{} 10 \\
\midrule
Capture: accepted / judged answers & 249/256 & 246/256 & (same) & 242/256 & 247/256 & (same) \\
Capture: generated tokens per accepted target & 3,568 & 2,057 & (same) & 3,533 & 2,023 & (same) \\
Training: supervised domain tokens (M) & 2.67 & 1.49 & 2.49 & 2.57 & 1.50 & 2.49 \\
Training: update time per seed (s) & 1,894 & 1,357 & 2,245 & 1,664 & 1,224 & 2,076 \\
Inference: median tokens per composition answer & 4,744 & 8,837 & 6,526 & 5,349 & 8,700 & 6,710 \\
Inference: mean tokens per composition answer & 6,025 & 9,198 & 7,071 & 6,264 & 9,073 & 7,658 \\
Inference: tokens per composition success & 10,206 & 26,491 & 16,556 & 13,666 & 29,032 & 19,346 \\
\bottomrule
\end{tabular}
\end{table}

\paragraph{Measured.}
Against the sheet-only policy of the same bank pair, the worked-solution policy needs 42\% fewer generated tokens per accepted target at capture and 28\% less update time for six passes (bank pair~1; 43\% and 26\% for bank pair~2), with nearly the same acceptance rate.
At inference, its models generate 53\% more tokens per composition answer (means of 9,198 against 6,025) and 2.6 times as many per composition success under the value signature (26,491 against 10,206); with bank pair~2 the factors are 1.45 and 2.1.
Ten worked-solution passes reduce the inference cost per composition success under the value signature (16,556 and 19,346 tokens) but take more update time than six sheet-only passes (2,245 against 1,894 seconds and 2,076 against 1,664) and remain less accurate on composition (Table~\ref{tab:banks}).
\textbf{Across the seed-aggregated bank-by-dose conditions in Table~\ref{tab:costs}, the worked-solution policy has neither a higher composition success rate nor fewer generated tokens per value-signature success than the sheet-only policy of its bank pair.}

\paragraph{Estimated.}
To compare the savings with the extra inference, we convert update time into generated-token equivalents at the measured generation throughput of the same GPU (1,191 tokens per second in the evaluation job of the second bank pair, with 64 concurrent requests).
On this conversion, the worked-solution policy of bank pair~1 saves 382,000 capture tokens and about 639,000 token equivalents of update time, and spends about 3,170 more tokens per composition query.
\textbf{Under this conversion, after roughly 320 composition queries the worked-solution policy has used more compute in total, while it has fewer value-signature successes} (about 310 queries for bank pair~2, computed the same way).
This estimate assumes that capture, training and inference compute are interchangeable at that rate; it ignores prefill and the cost of producing the worked solutions (a program here, but a solver or a stronger model in practice), and it holds only for this model, task mix and hardware.

\section{Interpretation, scope and limitations}
\label{sec:limits}

\paragraph{What the comparison shows.}
Two capture contexts that a practice-accuracy criterion would treat as equivalent produced models that composed the new rules differently, and the context that was cheaper to capture and train was the one that used more generated tokens per composition success under the value signature at inference.
The surrogate of Section~\ref{sec:account} offers one account of why the practice metric misses this: in the surrogate, practice accuracy measures progress along the direction that the targets set, not the direction itself.
The evaluation that exposed the difference needs no further training: unrequested-operation items with exact keys, answered without context (48 fresh composition items per state here).

\paragraph{Equal forward rates, not equal mastery.}
Equal observed forward rates do not establish equal mastery of the rules: the forward comparisons are superiority tests that did not reject, with intervals of about $\pm10$ to $\pm19$ points.
The claim concerns what a practitioner observes: at the forward rates that a held-out evaluation reports, identical on the new-argument panel (87.5\% each) and within 3 points on the fresh items, composition differed by 24 to 35 points.
Where forward rates differ more, we say so (bank pair~2 at six passes: 8.3 points; the answer-only policy: 15.3 points).
The trajectory view addresses a stronger form of the concern that the worked-solution models know the rules less well: along the sealed ladders, the worked-solution states lie lower at every forward rate that both policies reach (Figure~\ref{fig:curves}; exploratory).
It does not identify latent mastery.

\paragraph{How much of the cost goes with training dose.}
Worked-solution targets are shorter, so six passes supervise fewer domain tokens.
Trained for ten passes, close to the domain tokens of six sheet-only passes, the worked-solution policy gains composition in both bank pairs, and the pooled gap falls from 19.4 to 11.3 points.
With bank pair~1 a gap of 16.3 points remains and the pre-specified test holds; with bank pair~2 the gap is 6.2 points and the test fails, so for that pair a gap at matched dose is not established.
Ten passes also change replay, target visits, updates and learning-rate cycles, so neither result isolates one dose quantity.

\paragraph{Values of the training arguments.}
Memorized argument values do not explain the gap: at argument values that no training question used, both policies keep 87.5\% forward success, the composition gap is $+34.7$ points, and trained products are not easier than new ones (Section~\ref{sec:offsup}).
Mixed template-and-value mechanisms are not excluded.

\paragraph{Target banks and the judge.}
A second, independently captured bank pair gives the same direction at both doses, and every one of the five six-pass seed pairs on fresh items favors the sheet-only policy ($+14.6$ to $+33.3$ points); how much of the remaining variation comes from banks rather than training runs cannot be told from two banks (Section~\ref{sec:banks}).
The pre-specified measure is the program-checked signature, not the judge, and the judge agrees with it in direction on every test; because both read the same answers, this agreement does not validate the judge (Appendix~\ref{app:agreement}).

\paragraph{What the targets contain.}
``Unpracticed'' means not requested by the training questions.
A pattern scan of the targets found no inversion and a few passing mentions of a zero-current segment; a complete semantic reading was not done, because the reading model's content filter blocked it and we did not repeat it with another model.
This limits claims about what the models were never exposed to. The complete-policy comparison includes any differences in the generated targets; the present audit does not determine whether incidental exposure to compositional steps contributes to the observed gap.

\paragraph{Domain and model.}
The invented domain is what makes the measurement possible: the rules are new to the model, the keys are exact, and practiced and unpracticed uses are separated by construction.
Its cost is generality, which we do not claim.
The second model repeats the acquisition result, including at new argument values, but its unpracticed transfer was too weak for the policy comparison to be worth running.

\paragraph{Limitations.}
\begin{itemize}
\item \textbf{Mediator.} The worked solution changes target length, content, references and supervised tokens together, and we do not identify which change produces the gap. The surrogate of Section~\ref{sec:account} is one account; the absent-reference account and the package alternative are not excluded.
\item \textbf{Extensions are cells, not a grid.} New argument values were tested with bank pair~1 and six-pass states; bank pair~2 was tested on fresh items only.
\item \textbf{Variation.} Two target banks per main policy and one for the answer-only policy; two or three training seeds per bank. Item-bootstrap intervals do not contain variation between banks or training runs, and our data cannot separate these two sources. Comparisons across bank pairs and with the answer-only policy cross evaluation jobs that used the same request seeds and device type but different batching.
\item \textbf{Measures.} The signature checks numbers, not derivations; the official judge is a language model with sentinel checks, and the ten-pass verdicts of bank pair~1 come from a later batch than its six-pass verdicts. One numeric-parser repair was made after outputs existed (Appendix~\ref{app:prereg}); it changed no earlier signature.
\item \textbf{Scope.} One invented domain with two rules, the policy comparison on one model, one training recipe, and up to ten passes.
\item \textbf{Training records.} The run records of fourteen training passes (the third \genT{} and \instT{} seeds and the second \keptT{} seed) were not copied from the training storage during the study; the companion repository takes them from the author's archive of that storage and checks them against their training jobs under the trainer's rules, a file listing made on the storage before it was archived, and the evaluated checkpoints.
\end{itemize}

\paragraph{Future work, in order of what it would change.}
(1)~Several independently sampled banks per policy, each trained to matched dose with at least two seeds, to estimate how much the size of the cost varies between banks and training runs; the second bank pair on new argument values belongs here.
(2)~An edit test of the mediator: remove the references to the solution from worked-solution targets, against an edit-matched control at a fixed budget.
(3)~The policy comparison on a model whose unpracticed transfer is strong enough for adequate power, and in a second domain.
(4)~A masked human audit of a preselected answer panel and a semantic audit of the training targets.
(5)~For the answer-only policy, comparators captured in the same campaign.

\section{Conclusion}
\label{sec:conclusion}

Similar forward accuracy can conceal a substantial difference in the sequential reuse of newly learned rules.
In this controlled study, a worked solution made accepted self-generated targets shorter and reduced capture tokens and update time, yet the resulting models composed the rules less well.
At new argument values, both policies reached 87.5\% observed forward accuracy, while composition accuracy was 58.3\% with sheet-only targets and 23.6\% with worked-solution targets.
The gap was also confirmed prospectively on fresh items and repeated in direction with independently sampled targets.

Longer training reduced the difference.
At approximately matched domain-token dose, the original bank pair retained a 16.3-point gap, while the second pair's 6.2-point gap did not meet its pre-specified test.
Training amount therefore explains part of the observed separation; the remaining gap at matched dose is uncertain.
At six passes per policy, the worked-solution models used 2.6 times as many generated tokens per composition success under the value signature (2.1 times with the second bank pair), despite their lower capture-token and update-time requirements.

The practical lesson is to evaluate capture contexts by the uses they teach, not just the targets they make easier to produce.
For self-generated rule learning, test reuse beyond the practiced operation and count the inference cost of that reuse before selecting a context on forward accuracy alone.

\section*{Statements}
\label{sec:statement}
\addcontentsline{toc}{section}{Statements}

\paragraph{Author contribution and AI use.} Jiantong Xu conceived the research program, its ideas and the questions of this study, designed the experiments, set their methodological direction, made the scientific decisions as the work developed, and is responsible for all content. AI systems assisted under the author's direction: Claude (Anthropic; mainly Claude Opus 5.5 in Claude Code) and Codex (OpenAI) in experimental implementation and execution; Claude, Codex and ChatGPT (OpenAI) in scientific discussion and critical review of the design, analyses and manuscript; and Claude and ChatGPT in the analysis, the literature search and the drafting and revision of the manuscript. Claude Opus 5.5 also made the automated judgments described in Section~\ref{sec:setting}. The Kestrel items, keys and worked solutions are generated by the study's programs; the rule sheets and rubrics were written with AI assistance, and the battery-domain teaching texts of Appendix~\ref{app:battery} come from an earlier question bank of the project that was written with AI assistance and partly adapted from outputs of DeepSeek and Monica.

\paragraph{Funding.} This research was self-funded and received no external funding or other support.

\paragraph{Competing interests.} The author declares no competing interests.

\paragraph{Compute.} All training and inference used one NVIDIA A40 (48 GB) at a time. The service queue records 34 GPU hours of task time for the project's two studies of this question, including the familiar-domain study (Appendix~\ref{app:battery}) and a second-model candidate that failed its gate; billed machine time, including setup, was at least 36 hours. The new-argument evaluation, the likelihood scoring, the second bank pair and the answer-only policy added 6.9 hours of task time (7.25 hours billed), and the dose extension added 3.0 hours of task time (3.1 hours billed).

\paragraph{Availability.} The companion repository \href{https://github.com/JetXu-LLM/kestrel-capture-context}{\nolinkurl{github.com/JetXu-LLM/kestrel-capture-context}} (release v1.0.0) contains the item generator and all item sets, the protocols and amendments with a record of when each was first written, the training jobs and target selections, every model answer and judgment, the analysis code and derived results, and a script that recomputes on a CPU, from the saved answers and verdicts, every success rate, contrast, interval and test outcome in Sections~\ref{sec:results} and~\ref{sec:tests} and Appendix~\ref{app:variation}, every plotted value, and the answer lengths and outcome types of Sections~\ref{sec:diagnostics} and~\ref{sec:costs}; the statistics of the training targets and the capture and training costs come from the study's own analysis scripts, which the repository also contains. Every trained adapter, including the intermediate passes and the controls, is a release asset of the repository with its configuration, base-model revision and checksums. Code and adapters are released under the Apache License~2.0, and data, materials and documentation under CC~BY~4.0.

\bibliographystyle{plainnat}
\bibliography{references}

\begin{thebibliography}{29}
\providecommand{\natexlab}[1]{#1}
\providecommand{\url}[1]{\texttt{#1}}
\expandafter\ifx\csname urlstyle\endcsname\relax
  \providecommand{\doi}[1]{doi: #1}\else
  \providecommand{\doi}{doi: \begingroup \urlstyle{rm}\Url}\fi

\bibitem[Aky{\"u}rek et~al.(2024)Aky{\"u}rek, Aky{\"u}rek, Choshen, Wijaya, and
  Andreas]{akyurek2024deductive}
Afra~Feyza Aky{\"u}rek, Ekin Aky{\"u}rek, Leshem Choshen, Derry Wijaya, and
  Jacob Andreas.
\newblock Deductive closure training of language models for coherence,
  accuracy, and updatability.
\newblock \emph{arXiv preprint arXiv:2401.08574}, 2024.
\newblock Version 2.

\bibitem[Allen-Zhu and Li(2023)]{allenzhu2023physics31}
Zeyuan Allen-Zhu and Yuanzhi Li.
\newblock Physics of language models: Part 3.1, knowledge storage and
  extraction.
\newblock \emph{arXiv preprint arXiv:2309.14316}, 2023.
\newblock Version 3.

\bibitem[D'Amour et~al.(2020)D'Amour, Heller, Moldovan, Adlam, Alipanahi,
  Beutel, Chen, Deaton, Eisenstein, et~al.]{damour2020underspecification}
Alexander D'Amour, Katherine Heller, Dan Moldovan, Ben Adlam, Babak Alipanahi,
  Alex Beutel, Christina Chen, Jonathan Deaton, Jacob Eisenstein, et~al.
\newblock Underspecification presents challenges for credibility in modern
  machine learning.
\newblock \emph{arXiv preprint arXiv:2011.03395}, 2020.
\newblock Version 2.

\bibitem[Geirhos et~al.(2020)Geirhos, Jacobsen, Michaelis, Zemel, Brendel,
  Bethge, and Wichmann]{geirhos2020shortcut}
Robert Geirhos, J{\"o}rn-Henrik Jacobsen, Claudio Michaelis, Richard Zemel,
  Wieland Brendel, Matthias Bethge, and Felix~A. Wichmann.
\newblock Shortcut learning in deep neural networks.
\newblock \emph{arXiv preprint arXiv:2004.07780}, 2020.
\newblock Version 5.

\bibitem[Gekhman et~al.(2024)Gekhman, Yona, Aharoni, Eyal, Feder, Reichart, and
  Herzig]{gekhman2024newknowledge}
Zorik Gekhman, Gal Yona, Roee Aharoni, Matan Eyal, Amir Feder, Roi Reichart,
  and Jonathan Herzig.
\newblock Does fine-tuning {LLMs} on new knowledge encourage hallucinations?
\newblock In \emph{Proceedings of the 2024 Conference on Empirical Methods in
  Natural Language Processing}, 2024.
\newblock arXiv:2405.05904.

\bibitem[Hayou et~al.(2024)Hayou, Ghosh, and Yu]{hayou2024loraplus}
Soufiane Hayou, Nikhil Ghosh, and Bin Yu.
\newblock {LoRA+}: Efficient low rank adaptation of large models.
\newblock \emph{arXiv preprint arXiv:2402.12354}, 2024.
\newblock Version 2.

\bibitem[Hu et~al.(2022)Hu, Shen, Wallis, Allen-Zhu, Li, Wang, Wang, and
  Chen]{hu2022lora}
Edward~J. Hu, Yelong Shen, Phillip Wallis, Zeyuan Allen-Zhu, Yuanzhi Li, Shean
  Wang, Lu~Wang, and Weizhu Chen.
\newblock {LoRA}: Low-rank adaptation of large language models.
\newblock In \emph{International Conference on Learning Representations}, 2022.
\newblock arXiv:2106.09685.

\bibitem[Jiang et~al.(2026)Jiang, Xie, Jiang, and Zhang]{jiang2026skald}
Yubo Jiang, Fengying Xie, Zhiguo Jiang, and Haopeng Zhang.
\newblock Distill skills into weights, not prompts: Abstract skills as
  privileged signals for on-policy self-distillation.
\newblock \emph{arXiv preprint arXiv:2608.09826}, 2026.
\newblock Version 1.

\bibitem[Kim et~al.(2026)Kim, Luo, Kim, Lee, Kim, Jeon, Li, and
  Yang]{kim2026selfdistillation}
Jeonghye Kim, Xufang Luo, Minbeom Kim, Sangmook Lee, Dohyung Kim, Jiwon Jeon,
  Dongsheng Li, and Yuqing Yang.
\newblock Why does self-distillation (sometimes) degrade the reasoning
  capability of {LLMs}?
\newblock \emph{arXiv preprint arXiv:2603.24472}, 2026.
\newblock Version 4.

\bibitem[Lee et~al.(2026)Lee, Lee, Son, Lee, Han, Eo, and
  Lim]{lee2026answerconditioned}
Jungseob Lee, Seungyoon Lee, Suhyune Son, Dongyub~Jude Lee, Sungbin Han,
  Sugyeong Eo, and Heuiseok Lim.
\newblock Answer-conditioned chains of thought degrade verifiable-reasoning
  distillation in large language models.
\newblock \emph{arXiv preprint arXiv:2607.14552}, 2026.
\newblock Version 1.

\bibitem[Liu et~al.(2026{\natexlab{a}})Liu, Zhang, Yang, He, Wang, Xuan, Jin,
  and Diab]{liu2026mixsd}
Jiarui Liu, Lechen Zhang, Yongjin Yang, Yinghui He, Yingheng Wang, Weihao Xuan,
  Zhijing Jin, and Mona~T. Diab.
\newblock {MixSD}: Mixed contextual self-distillation for knowledge injection.
\newblock \emph{arXiv preprint arXiv:2605.16865}, 2026{\natexlab{a}}.
\newblock Version 3.

\bibitem[Liu et~al.(2026{\natexlab{b}})Liu, Dong, Li, Chen, Cao, Wang, Wang,
  Miao, Zhang, Wang, and Gao]{liu2026hpse}
Tianci Liu, Zihan Dong, Tianchun Li, Yi-Chung Chen, Qiming Cao, Xingchen Wang,
  Shiyang Wang, Zichen Miao, Linjun Zhang, Haoyu Wang, and Jing Gao.
\newblock Hybrid-policy self-editing for composable unstructured knowledge
  editing.
\newblock \emph{arXiv preprint arXiv:2608.11660}, 2026{\natexlab{b}}.
\newblock Version 1.

\bibitem[Lopez-Paz et~al.(2015)Lopez-Paz, Bottou, Sch{\"o}lkopf, and
  Vapnik]{lopezpaz2015unifying}
David Lopez-Paz, L{\'e}on Bottou, Bernhard Sch{\"o}lkopf, and Vladimir Vapnik.
\newblock Unifying distillation and privileged information.
\newblock \emph{arXiv preprint arXiv:1511.03643}, 2015.
\newblock Version 3.

\bibitem[Padmanabhan et~al.(2023)Padmanabhan, Onoe, Zhang, Durrett, and
  Choi]{padmanabhan2023propagating}
Shankar Padmanabhan, Yasumasa Onoe, Michael J.~Q. Zhang, Greg Durrett, and
  Eunsol Choi.
\newblock Propagating knowledge updates to {LMs} through distillation.
\newblock In \emph{Advances in Neural Information Processing Systems}, 2023.
\newblock arXiv:2306.09306.

\bibitem[Park et~al.(2025)Park, Zhang, and Tanaka]{park2025newnews}
Core~Francisco Park, Zechen Zhang, and Hidenori Tanaka.
\newblock New news: System-2 fine-tuning for robust integration of new
  knowledge.
\newblock \emph{arXiv preprint arXiv:2505.01812}, 2025.
\newblock Version 3.

\bibitem[Penaloza et~al.(2026)Penaloza, Vattikonda, Gontier, Lacoste, Charlin,
  and Caccia]{penaloza2026pidistill}
Emiliano Penaloza, Dheeraj Vattikonda, Nicolas Gontier, Alexandre Lacoste,
  Laurent Charlin, and Massimo Caccia.
\newblock Privileged information distillation for language models.
\newblock \emph{arXiv preprint arXiv:2602.04942}, 2026.
\newblock Version 3.

\bibitem[Press et~al.(2022)Press, Zhang, Min, Schmidt, Smith, and
  Lewis]{press2022compositionality}
Ofir Press, Muru Zhang, Sewon Min, Ludwig Schmidt, Noah~A. Smith, and Mike
  Lewis.
\newblock Measuring and narrowing the compositionality gap in language models.
\newblock \emph{arXiv preprint arXiv:2210.03350}, 2022.
\newblock Version 3.

\bibitem[Shenfeld et~al.(2026)Shenfeld, Damani, H{\"u}botter, and
  Agrawal]{shenfeld2026sdft}
Idan Shenfeld, Mehul Damani, Jonas H{\"u}botter, and Pulkit Agrawal.
\newblock Self-distillation enables continual learning.
\newblock \emph{arXiv preprint arXiv:2601.19897}, 2026.
\newblock Version 2.

\bibitem[Shrestha and Tessier(2026)]{shrestha2026rethinking}
Samyak Shrestha and Alexander Tessier.
\newblock Rethinking privileged information in on-policy self-distillation.
\newblock \emph{arXiv preprint arXiv:2608.18271}, 2026.
\newblock Version 1.

\bibitem[Snell et~al.(2022)Snell, Klein, and Zhong]{snell2022context}
Charlie Snell, Dan Klein, and Ruiqi Zhong.
\newblock Learning by distilling context.
\newblock \emph{arXiv preprint arXiv:2209.15189}, 2022.
\newblock Version 1.

\bibitem[Tang and Yang(2026)]{tang2026opd}
Yixuan Tang and Yi~Yang.
\newblock On-policy distillation teaches new skills but not new knowledge.
\newblock \emph{arXiv preprint arXiv:2610.09639}, 2026.
\newblock Version 1.

\bibitem[Tian et~al.(2026)Tian, Liu, Jiang, Dong, Li, Ding, Guo, Li, Hou, Wang,
  and Wang]{tian2026selfteacher}
Kanghui Tian, Siyuan Liu, Tianxiang Jiang, Shuai Dong, Yizhuo Li, Tian Ding,
  Yuan Guo, Songze Li, Haowen Hou, Congcong Wang, and Yi~Wang.
\newblock What should a self-teacher see? privileged context design for
  on-policy self-distillation.
\newblock \emph{arXiv preprint arXiv:2609.25623}, 2026.
\newblock Version 2.

\bibitem[Yang et~al.(2023)Yang, Lin, Zhou, and Wen]{yang2025rule}
Wenkai Yang, Yankai Lin, Jie Zhou, and Ji-Rong Wen.
\newblock Distilling rule-based knowledge into large language models.
\newblock \emph{arXiv preprint arXiv:2311.08883}, 2023.
\newblock Version 3.

\bibitem[Yuan et~al.(2025)Yuan, Chen, Zhang, Cui, Wang, You, Ding, Liu, Sun,
  and Peng]{yuan2025composing}
Lifan Yuan, Weize Chen, Yuchen Zhang, Ganqu Cui, Hanbin Wang, Ziming You, Ning
  Ding, Zhiyuan Liu, Maosong Sun, and Hao Peng.
\newblock From $f(x)$ and $g(x)$ to $f(g(x))$: {LLMs} learn new skills in {RL}
  by composing old ones.
\newblock \emph{arXiv preprint arXiv:2509.25123}, 2025.
\newblock Version 3.

\bibitem[Zelikman et~al.(2022)Zelikman, Wu, Mu, and Goodman]{zelikman2022star}
Eric Zelikman, Yuhuai Wu, Jesse Mu, and Noah~D. Goodman.
\newblock {STaR}: Bootstrapping reasoning with reasoning.
\newblock In \emph{Advances in Neural Information Processing Systems}, 2022.
\newblock arXiv:2203.14465.

\bibitem[Zhang et~al.(2026)Zhang, Chow, Fang, Liang, and Chua]{zhang2026ample}
XiuYu Zhang, Wei Chow, Junfeng Fang, Zhenkai Liang, and Tat-Seng Chua.
\newblock What does privileged information add to on-policy self-distillation?
\newblock \emph{arXiv preprint arXiv:2609.20612}, 2026.
\newblock Version 1.

\bibitem[Zhang et~al.(2024)Zhang, Khalifa, Logeswaran, Kim, Lee, Lee, and
  Wang]{zhang2024score}
Yunxiang Zhang, Muhammad Khalifa, Lajanugen Logeswaran, Jaekyeom Kim, Moontae
  Lee, Honglak Lee, and Lu~Wang.
\newblock Small language models need strong verifiers to self-correct
  reasoning.
\newblock In \emph{Findings of the Association for Computational Linguistics:
  ACL 2024}, pages 15637--15653, 2024.

\bibitem[Zhao et~al.(2026{\natexlab{a}})Zhao, Xie, Liu, Huang, Pang, Chen, and
  Grover]{zhao2026opsd}
Siyan Zhao, Zhihui Xie, Mengchen Liu, Jing Huang, Guan Pang, Feiyu Chen, and
  Aditya Grover.
\newblock Self-distilled reasoner: On-policy self-distillation for large
  language models.
\newblock \emph{arXiv preprint arXiv:2601.18734}, 2026{\natexlab{a}}.
\newblock Version 3.

\bibitem[Zhao et~al.(2026{\natexlab{b}})Zhao, Zhang, Xu, Wang, Caiyue, Zhao,
  and Li]{zhao2026psopsd}
Xuyang Zhao, Liting Zhang, Zichen Xu, Zhihu Wang, Xu~Caiyue, Shiwan Zhao, and
  Qicheng Li.
\newblock Is more privileged information better? from solution traces to
  problem-solving structure in self-distilled reasoning.
\newblock \emph{arXiv preprint arXiv:2608.01589}, 2026{\natexlab{b}}.
\newblock Version 1.

\end{thebibliography}

\appendix
\section{Acquisition, controls and side effects}
\label{app:acquisition}

\begin{table}[htbp]
\centering
\caption{\textbf{Six passes make the rules usable; at six passes the two policies have the same observed forward rate and different composition.} Official success (\%) of MiniCPM5-2B on the 124 sealed items without context, mean over seeds (number in brackets). $^{s}$: value-signature rates of states without an official reading (the protocol judged an arm's A and B items only once one of its states reached 20\% forward signature success, and assigned \instT{} after four passes to the signature).}
\label{tab:dose}
\small
\begin{tabular}{lcccc}
\toprule
State & Passes & Forward A (48) & Inversion (24) & Composition (24) \\
\midrule
Base model, no sheet [2 draws] & -- & 0.0 & 0.0 & 0.0 \\
\genT{} [3] & 2 & 8.3 & 2.8 & 2.8 \\
\genT{} [2] & 4 & 63.5 & 43.8 & 29.2 \\
\genT{} [2] & 6 & 82.3 & 72.9 & 68.8 \\
\instT{} [2] & 2 & 4.2 & 0.0 & 0.0 \\
\instT{} [2] & 4 & 63.5$^{s}$ & 22.9$^{s}$ & 14.6$^{s}$ \\
\instT{} [2] & 6 & 82.3 & 60.4 & 35.4 \\
\instT{} [2] & 10 & 89.6 & 56.2 & 50.0 \\
\keptT{} [2] & 6 & 0.0$^{s}$ & 0.0$^{s}$ & 6.2$^{s}$ \\
\bottomrule
\end{tabular}
\end{table}

The dose contrasts, recorded in the protocol as descriptive (without a margin), are six minus two passes $+74.0$\pp{} \ci{63.9}{83.3} and six minus four passes $+18.8$\pp{} \ci{6.3}{31.3}; a third six-pass \genT{} seed reaches 91.7\% forward signature success.
The pre-specified tests that the capture context changes forward acquisition did not reject: six \instT{} passes against six \genT{} passes $0.0$\pp{} \ci{-11.5}{11.5}, and ten \instT{} passes $-7.3$\pp{} \ci{-17.7}{2.1}.
On Qwen3-4B-Thinking-2507, six \genT{} passes raise forward signature success from 16.7\% (two passes) to 81.2\% (QT-1: $+64.6$\pp{} \ci{53.1}{76.0}); the test of unpracticed operations against the base model gives $+19.8$\pp{} \ci{11.5}{29.2} and misses its 20-point margin.
That \keptT{} fails to produce the rules does not show that nothing was learned internally.

\begin{table}[htbp]
\centering
\caption{\textbf{Ten worked-solution passes match the domain tokens of six sheet-only passes but also change every other dose quantity.} Training quantities of the evaluated states of bank pair~1 (MiniCPM5-2B), from the frozen training plans and the logged updates (the update counts of the third seed are planned counts; Section~\ref{sec:limits}).}
\label{tab:bundle}
\small
\setlength{\tabcolsep}{4pt}
\begin{tabular}{lrrrrr}
\toprule
State & Domain tokens & Replay tokens & Target visits & Updates (per seed) & LR cycles \\
\midrule
\genT{} 6 passes & 2,667,150 & 469,206 & 768 & 429 / 428 / 432 & 6 \\
\instT{} 6 passes & 1,494,891 & 469,206 & 762 & 284 / 292 / 281 & 6 \\
\instT{} 10 passes & 2,491,485 & 782,010 & 1,270 & 476 / 487 & 10 \\
\bottomrule
\end{tabular}
\end{table}

\paragraph{Abstention.}
On the 12 tested questions about a layer whose law is not given, the base model correctly says that the value is undetermined in 4 of 12 cases in each of two draws; after six \genT{} passes, both seeds answer 0 of 12 correctly ($-33.3$\pp{} \ci{-50.0}{-16.7}), usually by applying a learned or familiar law to the unknown layer.
This is a result on the tested panel, not a general calibration or safety evaluation.

\paragraph{Retention.}
On a 56-item retention panel, one six-pass \genT{} seed scores 18 of 24 professional answers against 22 for the base model, which reaches the pre-specified damage threshold; the other seed scores 21. The panel cannot establish general non-inferiority.

\paragraph{Declarative training first.}
Two passes on the model's own declarative explanations of the rules, followed by two passes of use practice, give 47.9\% forward success, below four passes of use practice ($-15.6$\pp{} \ci{-29.2}{-1.0}); about 37\% of the accepted declarative texts contained a wrong explanation, so this is not a test of correct declarative teaching.

\section{Sources of variation}
\label{app:variation}

The item-bootstrap intervals of the main text describe variation over items given the trained states.
This appendix shows the variation between training runs and between target banks that those intervals do not contain.
Each \genT{} seed is paired with the \instT{} seed of the same index; the two policies used the same pass seeds, and the second bank pair used the first two training seeds of the first.
The bank-pair differences and the same-seed comparisons across banks are exploratory analyses that we defined after all results existed.

\begin{figure}[htbp]
\centering
\includegraphics[width=\linewidth]{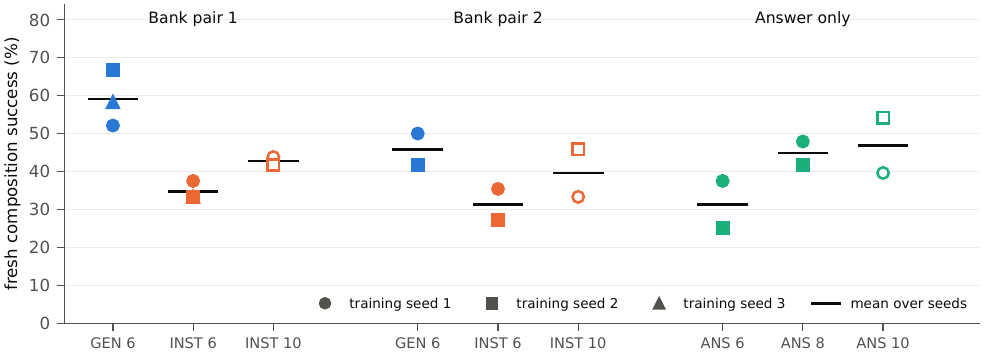}
\caption{\textbf{Training runs of one bank vary about as much as the two banks do.} Value-signature success on the 48 fresh composition items for every trained state, one mark per training seed (hollow: ten passes), with the mean over seeds as a bar.}
\label{fig:seeds}
\end{figure}

\begin{table}[htbp]
\centering
\caption{\textbf{Every six-pass seed pair favors the sheet-only policy; at matched dose three of four do, and the two bank pairs do not differ distinguishably.} Composition (value signature, 48 items per seed pair), sheet-only minus worked-solution policy in percentage points, with 95\% item-bootstrap intervals for the comparisons in the lower blocks.}
\label{tab:variation}
\small
\begin{tabular}{llr}
\toprule
Comparison & Seeds or contrast & Difference \\
\midrule
Six passes, fresh items, bank pair 1 & seeds 1, 2, 3 & $+14.6$, $+33.3$, $+25.0$ \\
Six passes, fresh items, bank pair 2 & seeds 1, 2 & $+14.6$, $+14.6$ \\
Six passes, new arguments, bank pair 1 & seeds 1, 2, 3 & $+29.2$, $+35.4$, $+39.6$ \\
Matched dose, fresh items, bank pair 1 & seeds 1, 2 & $+8.3$, $+25.0$ \\
Matched dose, fresh items, bank pair 2 & seeds 1, 2 & $+16.7$, $-4.2$ \\
\midrule
Same seed, two banks: \genT{} 6, bank 1 minus bank 2 & seed 1 & $+2.1$ \ci{-14.6}{18.8} \\
 & seed 2 & $+25.0$ \ci{6.3}{43.8} \\
Same seed, two banks: \instT{} 10, bank 1 minus bank 2 & seed 1 & $+10.4$ \ci{-6.3}{27.1} \\
 & seed 2 & $-4.2$ \ci{-22.9}{14.6} \\
\midrule
Gap of bank pair 1 minus gap of bank pair 2 & six passes & $+9.7$ \ci{-6.6}{27.1} \\
 & matched dose & $+10.1$ \ci{-5.6}{25.7} \\
 & six passes, official & $+9.0$ \ci{-7.6}{26.7} \\
 & matched dose, official & $+11.1$ \ci{-5.6}{27.1} \\
\bottomrule
\end{tabular}
\end{table}

\section{Additional diagnostics}
\label{app:diagnostics}

\paragraph{Outcome types under composition.}
On fresh two-current items, about two fifths of the answers of each six-pass policy carry the SOC that a constant capacity would give (Table~\ref{tab:errors}), although both apply the rate-dependent usable charge in 89\% of six-pass forward T2 answers.
The policies differ mainly in answers that match no known value.
These are descriptive categories of visible numbers; they do not identify a hidden route.

\begin{table}[htbp]
\centering
\caption{\textbf{Under composition the familiar law returns in both policies; the worked-solution policy adds answers that match no known value.} Value-signature outcome types on the fresh composition items of bank pair~1 (percent of answers). ``Familiar law'': a keyed quantity carries the value of a constant capacity or a familiar relaxation law; ``one current only'': the usable charge of one current used for both segments; ``partial'': some keyed quantities correct, none familiar; ``other'': numbers that match no keyed value. One \instT{} 10 two-current answer (2.1\%) had no final answer.}
\label{tab:errors}
\small
\begin{tabular}{llrrrrr}
\toprule
Items (answers) & State & Correct & Familiar law & One current only & Partial & Other \\
\midrule
Two-current (72/72/48) & \genT{} 6 & 38.9 & 40.3 & 2.8 & 0.0 & 18.1 \\
 & \instT{} 6 & 8.3 & 41.7 & 6.9 & 0.0 & 43.1 \\
 & \instT{} 10 & 16.7 & 60.4 & 2.1 & 0.0 & 18.8 \\
Rest-segment (72/72/48) & \genT{} 6 & 79.2 & 1.4 & -- & 4.2 & 15.3 \\
 & \instT{} 6 & 61.1 & 2.8 & -- & 22.2 & 13.9 \\
 & \instT{} 10 & 68.8 & 2.1 & -- & 18.8 & 10.4 \\
\bottomrule
\end{tabular}
\end{table}

\paragraph{The rules in solutions and on request.}
After six passes, both policies write the T1 factor $(1+h/\tau)^{-2}$ in 86--100\% of fresh forward T1 answers (2--14\% after two passes; \keptT{} 0\%).
Asked directly for the rules, the same models do not state them: in the nine answers of the three six-pass \genT{} seeds to the recall probes, none states either Kestrel relation.
We read this as elicitation behavior, in line with the format dependence of knowledge extraction \citep{allenzhu2023physics31}; it does not show that the knowledge is absent or inaccessible under every prompt.

\section{The Kestrel rule sheet}
\label{app:sheet}

The \genT{} capture context appends the following text to the system message (``Answer in English. Give the requested result and the scientific reason that supports it.''):

\begin{quote}\small
You know these relations. Use the relations that apply to the question. Write each relation that you use in full.
For a Kestrel cell with a source U, a series resistor R0, and a Varn layer in series, V = U - I R0 - v, where v is the Varn voltage. A constant-current segment starts at a current step and lasts until the next current step. At a time h after the start of a segment with current I, v = r v\_start + (1 - r) g I, with r = (1 + h/tau)\^{}(-2), where v\_start is v at the start of the segment. Here tau > 0 is the Varn time constant and g is the Varn gain. At an ideal current step, v is continuous. For a constant discharge current I > 0, the usable charge is Q\_use = Q\_nom (1 - k (I/I\_ref)\^{}(1/2)), with k > 0 and k (I/I\_ref)\^{}(1/2) < 1. During a constant-current discharge segment, the SOC decrease is the outward charge I h divided by Q\_use. One Ah is 3600 C.
\end{quote}

The \instT{} context adds a heading and a program-written worked solution of the question, which states the factor, the intermediate values and the final answer for that question only.
The \ansT{} context adds the heading ``Checked final answer of this question:'' and the final-answer line of that worked solution.

\section{Pre-specification record}
\label{app:prereg}

The study ran under two written protocols, of 26 October 2026 (the captures of bank pair~1, the first two training passes and their evaluation) and of 28 October (the training ladders and their sealed evaluation), and nine amendments to the second; each was written before the requests that it governs.
Amendments 1 and~2 (28 October) added the second model and the \instT{} dose ladder; amendment~3 (31 October) recorded the prospective tests TB-1 and TB-2; amendment~4 (1 November) recorded the semantic judging of the ten-pass composition answers and the new-argument tests OS-1 and OS-2; amendment~5 (2 November, before any request of that evaluation) fixed their reading rules; amendment~6 (2 November) recorded the predictions of the three binding forms; amendment~7 (5 November, before any of its requests) recorded the second bank pair, test RB-1 and the likelihood predictions; amendment~8 (5 November, 07:11~UTC) set two training seeds per new bank and added the answer-only policy with its gate and reading table; and amendment~9 (6 November, before any of its requests) recorded the dose extensions with test TB-2R and the answer-only second look.
The recorded campaign events place the write of amendment~3, including the TB-1 and TB-2 rules, at 00:52:36~UTC on 31~October 2026, the generation of the fresh items at 00:52:50 and the first evaluation request at 02:06:45; the heading of the amendment gives 01:30~UTC, which is not the recorded write time.
These dates come from the study's own records and execution logs, not from an external registry.
For each protocol and amendment, the companion repository gives its write time, every later change with its time, and the first queue or request record of every run that it governs (\texttt{protocol/CHRONOLOGY.md}).
Two kinds of departure from the records occurred.
First, the new-argument evaluation recorded on 1 November ran only on 5 November: no GPU was available in the data center that held the trained adapters during two bounded attempts, and a budget rule then stopped it until the author replaced the rule; its items and tests were unchanged, and its reading rules were revised (amendment~5) before any of its requests.
Second, one scoring repair followed the outputs: on 5~November, after the new-argument answers existed but before any reading of them, the value-signature parser was changed so that a product such as $35\times 10800$, which it had read as $35\times10^{10800}$ and which made a conversion overflow, becomes an infinite number that matches no key. The repair changed no earlier signature, and such misreadings can only make a number fail to match.

\begin{table}[htbp]
\centering
\caption{\textbf{Every pre-specified test, with its recorded probability and outcome, including the ones that failed.}}
\label{tab:prereg}
\small
\begin{tabular}{p{0.47\linewidth}llp{0.15\linewidth}}
\toprule
Test or prediction & Recorded & Probability & Outcome \\
\midrule
TI-1, TI-2 (the capture context changes forward acquisition) & A2 & 0.35, 0.20 & not held \\
QT-1 (second model, forward dose) & A1, A2 & 0.60 & held \\
QT-2 (second model, unpracticed operations, 20\pp{} margin) & A1, A2 & 0.50 & not held ($+19.8$) \\
\keptT{} six passes, forward signature at most 5\% (first; second seed) & protocol; A3 & 0.80; 0.85 & held (0\%) \\
TB-1 (fresh composition, same passes) & A3 & 0.60 & held \\
TB-2 (fresh composition, about the same domain tokens) & A3 & 0.40 & held \\
Official TB-2 at least $+10$\pp{} & A4 & 0.65 & held ($+18.4$) \\
OS-1 (new arguments, 30\pp{} margin) & A4 & 0.75 & held ($+87.5$) \\
OS-2 (new arguments, 10\pp{} margin) & A4 & 0.45 & held ($+34.7$) \\
RB-1 (second bank pair; two seeds per arm, replacing the three-seed version of A7, 0.60) & A8 & 0.55 & held ($+14.6$) \\
Answer-only policy, forward within 15\pp{} of \genT{} & A8 & 0.75 & not held ($-15.3$) \\
TB-2R (second bank pair, ten worked-solution passes) & A9 & 0.30 & not held ($+6.2$) \\
Answer-only policy after eight passes, forward gate & A9 & 0.80 & not held ($-17.4$) \\
Answer-only reading at the second look (ten passes) & A9 & 0.05 (graded) & graded \\
Familiar battery domain: T-D, T-S, T-V (Appendix~\ref{app:battery}) & earlier study & 0.65, 0.55, 0.45 & not held \\
Predictions of Table~\ref{tab:predictions} & A6 to A9 & 0.30--0.90 & see that table \\
\bottomrule
\end{tabular}
\end{table}

\section{Agreement of the two measures}
\label{app:agreement}

Table~\ref{tab:agreement} compares the value signature with the official endpoint on every answer of the study's second phase that has a semantic assessment.
The new-argument rows exclude the 24 forward answers of the base model (all failures on both measures); the row for the later banks covers every state of the second bank pair (six and ten passes) and of the answer-only policy (six, eight and ten passes).
Answers that failed the automatic checks of the endpoint, and judge dispositions without an assessment (for example, without a final answer), are endpoint failures and are excluded here.
The two measures read the same final answers, so their agreement is not an independent replication.
In the column ``Sig.\ T, off.\ fail'', the final answer contains every Kestrel value but the judge rejected its conclusion or support or found a contradiction.
On MiniCPM5-2B, all 13 sealed inversion answers in the last column belong to two items whose Kestrel value the signature classifies as ambiguous (counted as not T by the pre-specified rule), and for 8 of the 9 forward answers in that column the signature found, for at least one keyed quantity, the value of a familiar law and no Kestrel value.

\begin{table}[htbp]
\centering
\caption{\textbf{The two measures agree on 92.5--98.6\% of assessed answers per population.} Value signature against official endpoint, answers with a semantic verdict.}
\label{tab:agreement}
\footnotesize
\setlength{\tabcolsep}{4pt}
\begin{tabular}{llrrrr}
\toprule
Model & Items & Answers & Agree (\%) & Sig.\ T, off.\ fail & Sig.\ not T, off.\ success \\
\midrule
MiniCPM5-2B & sealed A & 479 & 94.8 & 16 & 9 \\
MiniCPM5-2B & sealed B-inv & 239 & 92.5 & 5 & 13 \\
MiniCPM5-2B & sealed B-comp & 240 & 93.8 & 15 & 0 \\
MiniCPM5-2B & fresh B-comp & 383 & 96.9 & 12 & 0 \\
MiniCPM5-2B & new-argument A & 72 & 98.6 & 1 & 0 \\
MiniCPM5-2B & new-argument B-comp & 288 & 93.8 & 14 & 4 \\
MiniCPM5-2B & fresh B-comp (later banks) & 572 & 95.6 & 25 & 0 \\
Qwen3-4B-Thinking & sealed A & 144 & 96.5 & 5 & 0 \\
Qwen3-4B-Thinking & sealed B-inv & 72 & 98.6 & 1 & 0 \\
Qwen3-4B-Thinking & sealed B-comp & 72 & 94.4 & 4 & 0 \\
\bottomrule
\end{tabular}
\end{table}

\section{A familiar-domain comparison that did not survive a dose control}
\label{app:battery}

The question of this paper arose in an earlier study on familiar battery questions (256 training questions; the same model and recipe).
There, captures with a general relation sheet beat captures with a worked solution on 48 sealed closed-book problems by $+21.5$\pp{}.
A recalibration of the judge reduced the gap to $+17.4$\pp{}, and a run that trained the worked-solution arm to about the same cumulative domain tokens left $+10.4$\pp{} \ci{-0.7}{21.5}, which did not meet its pre-specified criterion.
Two further pre-specified tests (targets with fewer uncertainty markers, and a context with only the checked final answer) also did not hold.
The primary claim for the familiar domain was not confirmed, and the residual difference is unresolved.
Unlike the Kestrel items, these problems use relations that the base model already knows, so they measure improved use of existing knowledge rather than the use of new rules.

\section{Literature reading record}
\label{app:literature}

We read the full texts (or, where marked, the abstract or named parts) of the closest prior work, at these versions: \citet{shenfeld2026sdft} (2601.19897v2, Sections~1 and~4 and Appendix~A.2); \citet{kim2026selfdistillation} (2603.24472v4, Sections~1--7); \citet{lee2026answerconditioned} (2607.14552v1, abstract, Sections~1, 3, 4.1, 6 and~7, Appendices~A and~G); \citet{zhang2024score} (Findings of ACL 2024, abstract, the opening of Section~3 and Section~3.1); \citet{zhao2026opsd} (2601.18734v3, abstract and Sections~3.2 and~4.2); \citet{tian2026selfteacher} (2609.25623v2, Sections~1--5, including Section~4.3); \citet{zhang2026ample} (2609.20612v1, abstract, Sections~1, 2 and~3.1 and the opening of Section~3.2); \citet{penaloza2026pidistill} (2602.04942v3, abstract, Sections~1, 3, 4.1--4.3 and~5--10, Appendix~D.1); \citet{shrestha2026rethinking} (2608.18271v1, abstract, Sections~1--2 and~4, Appendices~A.1, A.2 and~A.4); \citet{liu2026hpse} (2608.11660v1, abstract, introduction and method); \citet{liu2026mixsd} (2605.16865v3, Sections~3--6); \citet{park2025newnews} (2505.01812v3, Sections~3--5); \citet{yang2025rule} (2311.08883v3, Sections~4--5.1); \citet{yuan2025composing} (2509.25123v3, Sections~3--4.3); \citet{tang2026opd} (2610.09639v1, Sections~3--4.1); \citet{padmanabhan2023propagating} (2306.09306v2, Sections~1--4 and~8); \citet{snell2022context} (2209.15189v1, abstract and Section~3); \citet{zelikman2022star} (2203.14465v2, Sections~3.1--3.2); \citet{lopezpaz2015unifying} (1511.03643v3, abstract and the openings of Sections~3 and~4); \citet{damour2020underspecification} (2011.03395v2, abstract and Sections~2.1--2.2); \citet{geirhos2020shortcut} (2004.07780v5, abstract); \citet{press2022compositionality} (2210.03350v3, abstract); \citet{allenzhu2023physics31} (2309.14316v3, abstract); \citet{gekhman2024newknowledge} (2405.05904v3, abstract); \citet{akyurek2024deductive} (2401.08574v2, abstract); \citet{zhao2026psopsd} (2608.01589v1, abstract and an earlier full reading); \citet{jiang2026skald} (2608.09826v1, abstract and main results).
Focused searches covered offline rejection-sampling fine-tuning with hint-conditioned generation; comparisons of general principles and worked examples in context distillation or fine-tuning; references to an absent reference solution after self-distillation; and installing invented rules or operations and testing untrained operations.
The search was not exhaustive; arXiv search coverage is incomplete, and several on-policy papers were read at the abstract level only.

\end{document}